\documentclass[twocolumn]{aastex631}

\newcommand{\varv}{\ensuremath{v}}
\usepackage[dvipsnames]{xcolor}
\usepackage{booktabs}
\usepackage{graphicx}
\usepackage{array} 
\usepackage{tablefootnote}
\usepackage{amsmath}
\usepackage{txfonts}
\usepackage{hyperref}
\usepackage{balance}

\received{April 10, 2026}
\revised{May 29, 2026}
\accepted{June 2, 2026}

\shorttitle{Roasting Marshmallows survey: MASCARA-1b}
\shortauthors{Kanumalla et al.}

\begin{document}


\title{The Roasting Marshmallows Program with IGRINS on Gemini South V: Atmosphere of MASCARA-1\,b is Enriched in Refractory Elements}

\correspondingauthor{Krishna Kanumalla}
\email{krishna.kanumalla@asu.edu}

\author[0009-0005-4890-3326]{Krishna Kanumalla}
\affiliation{School of Earth and Space Exploration, Arizona State University, 781 Terrace Mall, Tempe, AZ, 85287, USA}

\author[0000-0002-2338-476X]{Michael R. Line}
\affiliation{School of Earth and Space Exploration, Arizona State University, 781 Terrace Mall, Tempe, AZ, 85287, USA}

\author[0009-0000-3360-6197]{Martina Chiarella}
\affiliation{Department of Physics, University of Turin, Via Pietro Giuria 1, I-10125, Turin, Italy}

\author[0000-0002-7704-0153]{Matteo Brogi}
\affiliation{Department of Physics, University of Turin, Via Pietro Giuria 1, I-10125, Turin, Italy}
\affiliation{INAF - Osservatorio Astrofisico di Torino, Via Osservatorio 20, I-10025 Pino Torinese, Italy}

\author[0000-0002-9946-5259]{Peter C. B. Smith}
\affiliation{School of Earth and Space Exploration, Arizona State University, 781 Terrace Mall, Tempe, AZ, 85287, USA}

\author[0000-0002-9142-6378]{Jorge A. Sanchez}
\affiliation{School of Earth and Space Exploration, Arizona State University, 781 Terrace Mall, Tempe, AZ, 85287, USA}

\author[0000-0003-1728-8269]{Yayaati Chachan}
\affiliation{Department of Astronomy and Astrophysics, University of California, Santa Cruz, CA 95064, USA}

\author[0000-0003-3667-8633]{Joshua Lothringer}
\affiliation{William H. Miller III Department of Physics \& Astronomy, Johns Hopkins University, 3400 N Charles Street, Baltimore, MD 21218, USA}
\affiliation{Space Telescope Science Institute, Baltimore, MD 21218, USA}

\author[0000-0003-3191-2486]{Joost P. Wardenier}
\affiliation{Weltraumforschung und Planetologie, Physikalisches Institut,University of Bern, Gesellschaftsstrasse 6, 3012 Bern, Switzerland}

\author[0000-0002-6980-052X]{Hayley Beltz}
\affiliation{Department of Physics and Astronomy, University of Kansas, Lawrence, KS, USA}

\author[0000-0002-0162-278X]{Carlos Saffe}
\affiliation{Instituto de Ciencias Astronómicas, de la Tierra y del Espacio (ICATE-CONICET), C.C 467, 5400 San Juan, Argentina}

\author[0000-0001-9796-2158]{Emily K. Deibert}
\affiliation{Department of Physics and Astronomy, University of Waterloo, 200 University Avenue West, Waterloo, Ontario N2L 3G1, Canada}
\affiliation{Waterloo Centre for Astrophysics, University of Waterloo, Waterloo, Ontario N2L 3G1, Canada}
\affiliation{International Gemini Observatory/NSF NOIRLab, Casilla 603, La Serena, Chile}

\author[0000-0003-4241-7413]{Megan Weiner Mansfield}
\affiliation{Department of Astronomy, University of Maryland, College Park, MD, USA}

\author[0000-0002-8573-805X]{Stefan Pelletier}
\affiliation{Observatoire de Genève, Département d’Astronomie, Université de Genève, Chemin Pegasi 51, 1290 Versoix, Switzerland}

\author[0000-0001-9521-6258]{Vivien Parmentier}
\affiliation{Universite Cote d’Azur, Av. Valrose, 06000 Nice, France}

\author[0009-0005-5145-5165]{Yeon-ho Choi}
\affiliation{Korea Astronomy and Space Science Institute, 776 Daedeokdae-ro, Yuseong-gu, Daejeon 34055, Republic of Korea}

\author[0000-0003-0815-8366]{Swaetha Ramkumar}
\affiliation{School of Earth and Space Exploration, Arizona State University, 781 Terrace Mall, Tempe, AZ, 85287, USA}

\author[0000-0002-2454-768X]{Arjun B. Savel}
\affiliation{Department of Astronomy, University of Maryland, College Park, MD, USA}

\author[0000-0003-0156-4564]{Luis Welbanks}
\affiliation{School of Earth and Space Exploration, Arizona State University, 781 Terrace Mall, Tempe, AZ, 85287, USA}

\author[0000-0003-4733-6532]{Jacob L.\ Bean}
\affiliation{Department of Astronomy \& Astrophysics, University of Chicago, Chicago, IL 60637}

\author[0000-0002-2513-4465]{Vatsal Panwar}
\affiliation{School of Physics \& Astronomy, University of Birmingham, Edgbaston, Birmingham B15 2TT, UK}

\author[0000-0002-9379-4895]{Tomás Azevedo Silva}
\affiliation{INAF - Osservatorio Astrofisico di Arcetri, Largo Enrico Fermi 5, I-50125 Firenze, Italy}

\author[0000-0002-1321-8856]{Lorenzo Pino}
\affiliation{INAF - Osservatorio Astrofisico di Arcetri, Largo Enrico Fermi 5, I-50125 Firenze, Italy}

\author[0000-0001-8877-0242]{Yuya Hayashi}
\affiliation{Department of Multi-Disciplinary Sciences, The University of Tokyo, 3-8-1 Komaba, Meguro, Tokyo 153-8902, Japan}

\author[0000-0001-7277-7175]{Dongwook Lim}
\affiliation{Center for Galaxy Evolution Research \& Department of Astronomy, Yonsei University, 50 Yonsei-ro, Seoul 03722, Republic of Korea}

\author[0000-0001-9352-0248]{Cicero X. Lu}
\affiliation{International Gemini Observatory/NSF NOIRLab, 670 N. A’ohoku Place, Hilo, Hawai’i, 96720, USA}

\author{Venu M. Kalari}
\affiliation{International Gemini Observatory/NSF NOIRLab, Casilla 603, La Serena, Chile}

\author[0000-0003-4603-556X]{Teo Mo\v{c}nik}
\affiliation{International Gemini Observatory/NSF NOIRLab, 670 N. A’ohoku Place, Hilo, Hawai’i, 96720, USA}

\author[0000-0002-6529-202X]{Mark G. Rawlings}
\affiliation{International Gemini Observatory/NSF NOIRLab, 670 N. A’ohoku Place, Hilo, Hawai’i, 96720, USA}

\author[0000-0002-0418-5335]{Heeyoung Oh}
\affiliation{Korea Astronomy and Space Science Institute, 776 Daedeokdae-ro, Yuseong-gu, Daejeon 34055, Republic of Korea}

\author[0000-0001-9716-5335]{Ruben J. Diaz}
\affiliation{International Gemini Observatory/NSF NOIRLab, 950 N. Cherry Ave., Tucson, AZ 85719, USA}
\affiliation{Universidad Nacional de C\'ordoba Laprida 854, C\'ordoba, X5000BGR, Argentina.}

\author[0000-0001-9773-3080]{Chan Park}
\affiliation{Korea Astronomy and Space Science Institute, 776 Daedeokdae-ro, Yuseong-gu, Daejeon 34055, Republic of Korea}

\author[0000-0003-0894-7824]{Jae-Joon Lee}
\affiliation{Korea Astronomy and Space Science Institute, 776 Daedeokdae-ro, Yuseong-gu, Daejeon 34055, Republic of Korea}

\author{Sanghyuk Kim}
\affiliation{Korea Astronomy and Space Science Institute, 776 Daedeokdae-ro, Yuseong-gu, Daejeon 34055, Republic of Korea}

\author{Ueejeong Jeong}
\affiliation{Korea Astronomy and Space Science Institute, 776 Daedeokdae-ro, Yuseong-gu, Daejeon 34055, Republic of Korea}
\affiliation{University of Texas at Austin, 2515 Speedway, Stop C1400, Austin, Texas 78712-1205, USA}

\author{Hye-In Lee}
\affiliation{Korea Astronomy and Space Science Institute, 776 Daedeokdae-ro, Yuseong-gu, Daejeon 34055, Republic of Korea}

\author[0000-0001-8012-5871]{Woojin Park}
\affiliation{Korea Astronomy and Space Science Institute, 776 Daedeokdae-ro, Yuseong-gu, Daejeon 34055, Republic of Korea}

\author{Youngsam Yu}
\affiliation{Korea Astronomy and Space Science Institute, 776 Daedeokdae-ro, Yuseong-gu, Daejeon 34055, Republic of Korea}

\author{Yunjong Kim}
\affiliation{Korea Astronomy and Space Science Institute, 776 Daedeokdae-ro, Yuseong-gu, Daejeon 34055, Republic of Korea}

\author{Moo-Young Chun}
\affiliation{Korea Astronomy and Space Science Institute, 776 Daedeokdae-ro, Yuseong-gu, Daejeon 34055, Republic of Korea}

\author{Jae Sok Oh}
\affiliation{Korea Astronomy and Space Science Institute, 776 Daedeokdae-ro, Yuseong-gu, Daejeon 34055, Republic of Korea}

\author{Sungho Lee}
\affiliation{Korea Astronomy and Space Science Institute, 776 Daedeokdae-ro, Yuseong-gu, Daejeon 34055, Republic of Korea}

\author{Jeong-Gyun Jang}
\affiliation{Korea Astronomy and Space Science Institute, 776 Daedeokdae-ro, Yuseong-gu, Daejeon 34055, Republic of Korea}

\author{Bi-Ho Jang}
\affiliation{Korea Astronomy and Space Science Institute, 776 Daedeokdae-ro, Yuseong-gu, Daejeon 34055, Republic of Korea}

\author{Hyeon Cheol Seong}
\affiliation{Korea Astronomy and Space Science Institute, 776 Daedeokdae-ro, Yuseong-gu, Daejeon 34055, Republic of Korea}

\author[0000-0001-9263-3275]{Hyun-Jeong Kim}
\affiliation{Korea Astronomy and Space Science Institute, 776 Daedeokdae-ro, Yuseong-gu, Daejeon 34055, Republic of Korea}

\author{Cynthia B. Brooks}
\affiliation{University of Texas at Austin, 2515 Speedway, Stop C1400, Austin, Texas 78712-1205, USA}

\author[0000-0001-7875-6391]{Gregory N. Mace}
\affiliation{University of Texas at Austin, 2515 Speedway, Stop C1400, Austin, Texas 78712-1205, USA}

\author{Hanshin Lee}
\affiliation{University of Texas at Austin, 2515 Speedway, Stop C1400, Austin, Texas 78712-1205, USA}

\author{John M. Good}
\affiliation{University of Texas at Austin, 2515 Speedway, Stop C1400, Austin, Texas 78712-1205, USA}

\author[0000-0003-3577-3540]{Daniel T. Jaffe}
\affiliation{University of Texas at Austin, 2515 Speedway, Stop C1400, Austin, Texas 78712-1205, USA}

\author{Kang-Min Kim}
\affiliation{Korea Astronomy and Space Science Institute, 776 Daedeokdae-ro, Yuseong-gu, Daejeon 34055, Republic of Korea}

\author{In-Soo Yuk}
\affiliation{Korea Astronomy and Space Science Institute, 776 Daedeokdae-ro, Yuseong-gu, Daejeon 34055, Republic of Korea}

\author[0000-0002-2013-1273]{Narae Hwang}
\affiliation{Korea Astronomy and Space Science Institute, 776 Daedeokdae-ro, Yuseong-gu, Daejeon 34055, Republic of Korea}

\author[0000-0002-6982-7722]{Byeong-Gon Park}
\affiliation{Korea Astronomy and Space Science Institute, 776 Daedeokdae-ro, Yuseong-gu, Daejeon 34055, Republic of Korea}

\author[0000-0003-4770-688X]{Hwihyun Kim}
\affiliation{International Gemini Observatory/NSF NOIRLab, 950 N. Cherry Ave., Tucson, AZ 85719, USA}

\author{Brian Chinn}
\affiliation{International Gemini Observatory/NSF NOIRLab, Casilla 603, La Serena, Chile}

\author{Francisco Ramos}
\affiliation{International Gemini Observatory/NSF NOIRLab, Casilla 603, La Serena, Chile}

\author{Pablo Prado}
\affiliation{International Gemini Observatory/NSF NOIRLab, Casilla 603, La Serena, Chile}

\author{John White}
\affiliation{International Gemini Observatory/NSF NOIRLab, 670 N. A’ohoku Place, Hilo, Hawai’i, 96720, USA}

\author{Andres Olivares}
\affiliation{International Gemini Observatory/NSF NOIRLab, Casilla 603, La Serena, Chile}

\author{Valentina Oyarzun}
\affiliation{International Gemini Observatory/NSF NOIRLab, Casilla 603, La Serena, Chile}

\author{Emma Kurz}
\affiliation{International Gemini Observatory/NSF NOIRLab, 670 N. A’ohoku Place, Hilo, Hawai’i, 96720, USA}
\author{Hawi Stecher}
\affiliation{International Gemini Observatory/NSF NOIRLab, 670 N. A’ohoku Place, Hilo, Hawai’i, 96720, USA}
\author{Carlos Quiroz}
\affiliation{International Gemini Observatory/NSF NOIRLab, Casilla 603, La Serena, Chile}

\author{Ignacio Arriagada}
\affiliation{International Gemini Observatory/NSF NOIRLab, Casilla 603, La Serena, Chile}

\author{Thomas L. Hayward}
\affiliation{International Gemini Observatory/NSF NOIRLab, Casilla 603, La Serena, Chile}

\author[0000-0002-2536-1633]{Hyewon Suh}
\affiliation{International Gemini Observatory/NSF NOIRLab, 670 N. A’ohoku Place, Hilo, Hawai’i, 96720, USA}
\author{Jen Miller}
\affiliation{International Gemini Observatory/NSF NOIRLab, 670 N. A’ohoku Place, Hilo, Hawai’i, 96720, USA}
\author{Siyi Xu}
\affiliation{International Gemini Observatory/NSF NOIRLab, 670 N. A’ohoku Place, Hilo, Hawai’i, 96720, USA}
\author[0000-0002-6822-2254]{Emanuele Paolo Farina}
\affiliation{International Gemini Observatory/NSF NOIRLab, 670 N. A’ohoku Place, Hilo, Hawai’i, 96720, USA}
\author{Charlie Figura}
\affiliation{International Gemini Observatory/NSF NOIRLab, 670 N. A’ohoku Place, Hilo, Hawai’i, 96720, USA}
\author{Andrew Stephens}
\affiliation{International Gemini Observatory/NSF NOIRLab, 670 N. A’ohoku Place, Hilo, Hawai’i, 96720, USA}
\author[0000-0002-5665-376X]{Bryan Miller}
\affiliation{International Gemini Observatory/NSF NOIRLab, Casilla 603, La Serena, Chile}

\author{Kathleen Labrie}
\affiliation{International Gemini Observatory/NSF NOIRLab, 670 N. A’ohoku Place, Hilo, Hawai’i, 96720, USA}

\author{Paul Hirst}
\affiliation{International Gemini Observatory/NSF NOIRLab, 670 N. A’ohoku Place, Hilo, Hawai’i, 96720, USA}

\author{Edo Tapia}
\affiliation{International Gemini Observatory/NSF NOIRLab, 670 N. A’ohoku Place, Hilo, Hawai’i, 96720, USA}

\author{Zachary Hartmann}
\affiliation{International Gemini Observatory/NSF NOIRLab, 670 N. A’ohoku Place, Hilo, Hawai’i, 96720, USA}




\begin{abstract}

Ultra-hot Jupiters (UHJs; T$_{\mathrm{eq}} \gtrsim 2000$ K) enable simultaneous detection of volatile (ice-forming) and refractory (rock-forming) species in planetary atmospheres, providing a powerful diagnostic of planet formation and atmospheric processing. We present a comprehensive high-resolution cross-correlation spectroscopy (HRCCS) analysis of the UHJ MASCARA-1\,b (T$_{\mathrm{eq}} \approx 2600$ K) using the IGRINS and IGRINS-2 spectrographs. We detect robust (SNR$>$4) signals from H$_2$O, CO, OH, Fe\,\textsc{i}, Mg\,\textsc{i}, Ca\,\textsc{i}, and Ti\,\textsc{i}, marking the most complete atmospheric inventory of MASCARA-1\,b to date. Using a chemically consistent atmospheric inference framework, we constrain elemental abundances to a typical precision of $\approx$ 0.2 dex, retrieving a solar atmospheric metallicity ([M/H]$_\odot$ $= 0.07^{+0.17}_{-0.13}$ $\approx 1.2\times$ solar), a C/O ratio (C/O $= 0.65^{+0.08}_{-0.08}$) consistent with solar value (C/O $=$ 0.59), an enhanced refractory abundance ([$\mathcal{R}$/H]$_\odot$ = $0.40^{+0.23}_{-0.17} \approx 2.5\times$ solar; $\approx 3.8\times$ stellar), and a moderately super-solar refractory-to-volatile ratio ([$\mathcal{R}/\mathcal{V}$]$_\odot =$  $0.36^{+0.11}_{-0.09}$ $\approx$ 2.3$\times$ solar). Comparison with formation models suggests that MASCARA-1\,b most likely accreted material between the soot–H$_2$O or H$_2$O–CO snowlines (at 68\% confidence). We additionally find stellar values for atmospheric Ti/Mg and Ca/Mg ratios (at 68\% confidence). The Mg/Fe is also found to be consistent with stellar value at 95\% confidence. Therefore, we do not find strong indication of nightside cold trapping in MASCARA-1~b. As homogeneous refractory-to-volatile measurements expand across the UHJ population, particularly with upcoming Extremely Large Telescopes, these diagnostics will enable statistically robust tests of emerging trends in giant planet formation and atmospheric evolution.

\end{abstract}


\keywords{: Exoplanet atmospheres (487) -- Exoplanet atmospheric composition (2021) -- Exoplanet atmospheric structure (2310) -- High resolution spectroscopy (2096) -- Infrared spectroscopy (2285)}

\section{Introduction} \label{sec:intro}

As the relics of formation and evolution in the protoplanetary disk, primary atmospheres of exoplanets provide direct insight into the physicochemical processes that shape planetary systems. In particular, measurements of  gas giant atmospheric composition have complemented and expanded on what can be learned from  our Solar System alone \citep[e.g.,][]{Dawson2018, Guillot23_review}, enabling inferences on formation locations, migration histories, and the relative roles of gas and solid accretion in protoplanetary disks. For close-in gas giants, high equilibrium temperatures place many key molecular and atomic species in the gas phase \citep[e.g.,][]{Hoeij18, gandhi2023_hrs_dynamics, co2_jwst, Smith24_W121b}, allowing their abundances to be probed spectroscopically with high precision.

A commonly used diagnostic of giant planet formation is the atmospheric carbon-to-oxygen ratio (C/O), which is expected to vary with formation location relative to major snow lines in the protoplanetary disk \citep[e.g.,][]{oberg2011, mordasini2016}. However, recent theoretical work has demonstrated that C/O alone is insufficient to uniquely constrain formation pathways, as similar C/O ratios can arise from markedly different combinations of gas accretion, solid accretion, and elemental sequestration \citep[e.g.,][]{Lothringer2021}. As a result, additional compositional diagnostics are required to robustly link observed atmospheric abundances to planet formation histories \citep[][]{Turrini2021, Chachan2023}.

Simultaneous detections of species containing volatile and refractory elements provide a powerful means of breaking this degeneracy \citep{Lothringer2021}, as the relative enrichment of refractory to volatile abundance encodes the balance between solid and gas accretion during planet formation \citep[e.g.,][]{Chachan2023}. Ultra-hot Jupiters (UHJs; T$_\mathrm{eq} \gtrsim$ 2000~K) are uniquely well suited to this approach as their extreme dayside temperatures prevent condensation of many refractory species, allowing both volatile molecules (e.g., CO, H$_2$O, OH) and refractory atoms (e.g., Fe\,\textsc{i}, Ti\,\textsc{i}, Mg\,\textsc{i}, Si\,\textsc{i}, Cr\,\textsc{i}, Ca\,\textsc{i}) to remain in the gas phase and produce detectable spectral signatures \citep[e.g.,][]{Lothringer2021, Kasper2023, Deibert24_W189b, pelletier_VO_HRS, Smith24_W121b}. Consequently, UHJs offer a rare opportunity to directly measure refractory-to-volatile ($\mathcal{R}/\mathcal{V}$) ratios in exoplanet atmospheres.

High-resolution spectroscopy has proven to be particularly beneficial  for probing the atmospheres of UHJs, as it enables the detection of individual atomic and molecular species through their resolved line profiles and Doppler shifts \citep{sne10, Bir18, Snellen2025_review}. In emission, this technique directly probes the dayside atmosphere and is sensitive to both atmospheric composition and thermal structure, the presence of temperature inversions, and 3D effects \citep[e.g.,][]{nugroho17, Line2021_W77Ab, Brogi2023, Beltz2021_3D_HD209}. Recent advances in high-resolution retrieval frameworks now allow this technique to be used not only for species detection, but also for quantitative constraints on elemental abundances and atmospheric structure \citep{BL19, Gibson20_W121b, pellet2021, gandhi2023_hrs_dynamics}.

\begin{figure*}
    \centering
    \includegraphics[width=\textwidth]{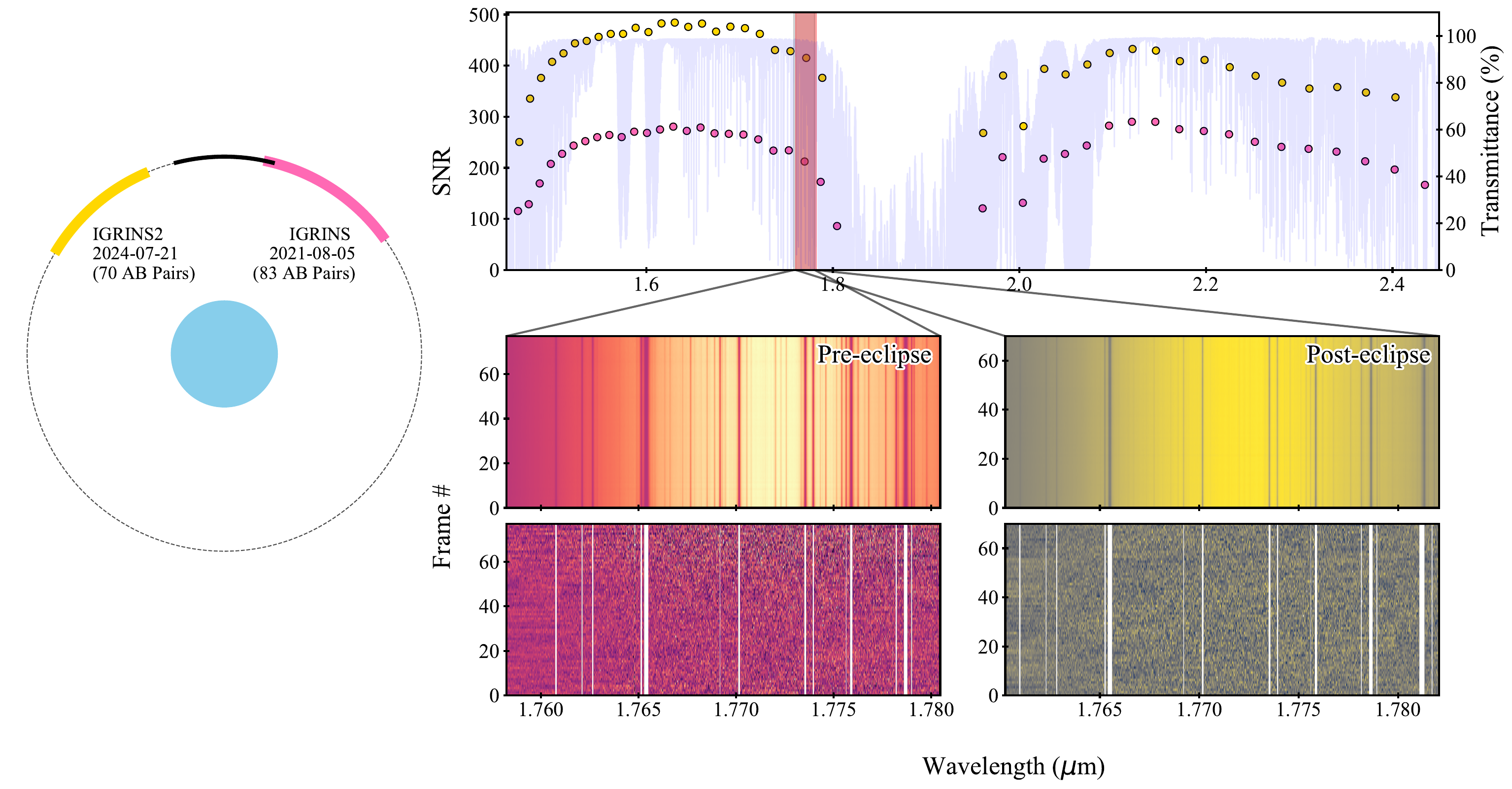}
    \caption{Summary of our observations and detrending process. Left inset shows the phase coverage of our observations in pre-secondary eclipse (pink) and post-secondary eclipse (yellow) geometries. The phase range obstructed by the secondary eclipse is shown as the thick black line between these geometries. In the top right panel, we show our median signal-to-noise ratios (SNRs) per order for pre-eclipse (pink) and post-eclipse (yellow) datasets. IGRINS and IGRINS-2 offer near-identical wavelength coverage in H and K bands. Due to better observing conditions, our post-eclipse dataset has a higher SNR than the pre-eclipse observations. In the bottom panels to the right, we show the raw data and post-SVD residual data (bottom) from an H-band order with heavy telluric contamination. The transmittance spectrum of Earth's atmosphere is shown in the background of top SNR panel. The white regions in the residual data are the masked wavelengths where there is high telluric contamination.}
    \label{fig:observations and SVD summarized}
\end{figure*}

MASCARA-1\,b (T$_\mathrm{eq}$ $\approx$ 2594 K, R$_\mathrm{p}$ $\approx$ 1.597 R$_\mathrm{Jup}$, M$_\mathrm{p}$ $\approx$ 3.7 M$_\mathrm{Jup}$) is an exceptionally favorable target for this type of analysis \citep[][Figure \ref{fig:survey}]{Talens2017}. Despite orbiting one of the brightest (m$_{K}$ = 7.7) known UHJ host stars, atmospheric characterization of MASCARA-1\,b in transmission has proven difficult due to its high surface gravity and the near-complete overlap between the planetary orbital velocity and the stellar Doppler shadow, resulting in non-detections or upper limits \citep[e.g.,][]{CasasayasBarris2022, Stangret22_HATP57b_KELT17b_M1b_W189b_etc}. In contrast, the planet’s extremely hot dayside and expected inverted temperature structure make it well suited for emission spectroscopy, where several studies have successfully detected individual volatile (H$_2$O, CO) and refractory (Fe\,\textsc{i}, Ti\,\textsc{i}, Cr\,\textsc{i}) species at optical and infrared wavelengths \citep{Scandariato23_M1b,Guo2024,Ramkumar2023}. In Appendix Table \ref{tab:detections}, we have compiled the list of detections and non-detections in MASCARA-1\,b's atmosphere from previous studies.

In this work, we present a unified atmospheric analysis of MASCARA-1\,b based on high-resolution emission spectroscopy that combines new IGRINS-2 system verification observations with complementary IGRINS observations from the \textit{Roasting Marshmallows} survey (Figure \ref{fig:survey}). We detect and constrain the abundances of a broad set of volatile and refractory species, yielding the most complete species inventory of MASCARA-1\,b to date. This comprehensive view enables direct comparisons between volatile and refractory abundances and places new constraints on the planet’s atmospheric composition and formation history. 

In Section \ref{sec:observations}, we describe our observations and steps taken to detrend our data. In Section \ref{sec:cc_and_detections}, we present the cross-correlation analysis and species detections. In Section \ref{sec:retrievals}, we apply a Bayesian analysis to retrieve the elemental composition of MASCARA-1\,b's atmosphere, deriving that it is enriched in refractory elements. We discuss the formation of MASCARA-1\,b in Section \ref{sec:formation}. We also discuss the inferences drawn from our study and previous Roasting Marshmallows survey results in Sections \ref{sec:UHJ_sample} and \ref{sec:coldtrapping} respectively. Finally, we summarize our conclusions in Section \ref{sec:conclusions}.

\begin{table*}[ht!]
\centering
\caption{Adopted System parameters}
\label{tab:parameters}
\begin{tabular}{llll}
\toprule
Parameter & Symbol & Value $\pm$ Error [unit] & Reference\\
\midrule
Stellar Effective Temperature & T$_\mathrm{eff}$ & 7490$\pm$150 K & \cite{Hooton2022} \\
Stellar logarithmic surface gravity & log$_\mathrm{10}$ \textit{g$_\star$} & 4.09$\pm$0.006 c.g.s & \cite{Hooton2022} \\
Stellar Radius & R$_\mathrm{s}$ & 2.082$^{+0.022}_{-0.024}$ R$_\odot$ & \cite{Hooton2022} \\
Stellar Mass & M$_\mathrm{s}$ & 1.9$^{+0.068}_{-0.063}$ M$_\odot$ & \cite{Hooton2022} \\
Stellar rotation velocity & $\varv$ sin\textit{i}$_*$ & 101.7$^{+3.5}_{-4.2}$ km/sec & \cite{Hooton2022} \\
& & & \\
& [C/H] & $-0.32\pm0.08$ & \cite{Saffe2021} \\
Stellar abundances (in log) & [Fe/H] & $-0.11\pm0.15$ & \cite{Saffe2021} \\
(relative to solar value) & [Mg/H] & $-0.33\pm0.24$ & \cite{Saffe2021} \\
& [Ca/H] & $0.13\pm0.10$ & \cite{Saffe2021} \\
& [Ti/H] & $-0.04\pm0.13$ & \cite{Saffe2021} \\
\midrule
Planetary radius & R$_\mathrm{pl}$ & 1.597$^{+0.018}_{-0.019}$ R$_\mathrm{Jup}$ & \cite{Hooton2022} \\
Planetary mass & M$_\mathrm{Jup}$ & 3.7$\pm$0.9 M$_\mathrm{Jup}$ & \cite{Hooton2022} \\
Orbital Period & Per & 2.14877381$\pm$0.00000088 days & \cite{Hooton2022} \\
Semi major axis & a & 0.040352$^{+0.000046}_{-0.000049}$ au & \cite{Hooton2022} \\
Transit Midpoint & T$_\mathrm{c}$ & 2458833.488151$\pm$0.000091 BJD$_\mathrm{TDB}$ & \cite{Hooton2022} \\
Planetary equilibrium temperature & T$_\mathrm{eq}$ & 2594$\pm$2 K & \cite{Hooton2022} \\
Systemic velocity & V$_\mathrm{sys}$ & 11.2$\pm$0.08 km/sec$^\mathrm{a}$ & \cite{Talens2017} \\
Planetary logarithmic surface gravity & log$_\mathrm{10}$ \textit{g} & 3.57$\pm$0.3 c.g.s & Derived \\
Planetary rotation velocity (solid body) & $\varv$ sin\textit{i}$_\mathrm{pl}$ = $\mathrm{\frac{2\pi R_\mathrm{pl}}{Per}}$ & 3.55$\pm$0.02 km/sec & Derived \\
Semi amplitude of planet's radial velocity & K$_\mathrm{p}$ = $\mathrm{\frac{2\pi a}{Per}}$ & 204.2$\pm$0.2 km/sec & Derived \\
\bottomrule
\end{tabular}
\begin{minipage}{\textwidth}
\footnotesize
$^\mathrm{a}$ Based on HERMES radial velocity data. However, \cite{Talens2017} analysis also reports a value of $\sim$ 8.5 km/sec based on SONG observations. Using ESPRESSO data, \cite{CasasayasBarris2022} also derive an estimate of $\sim$ 9.3 km/sec.
\end{minipage}
\end{table*}

\begin{figure*}
    \centering
    \includegraphics[width=\textwidth]{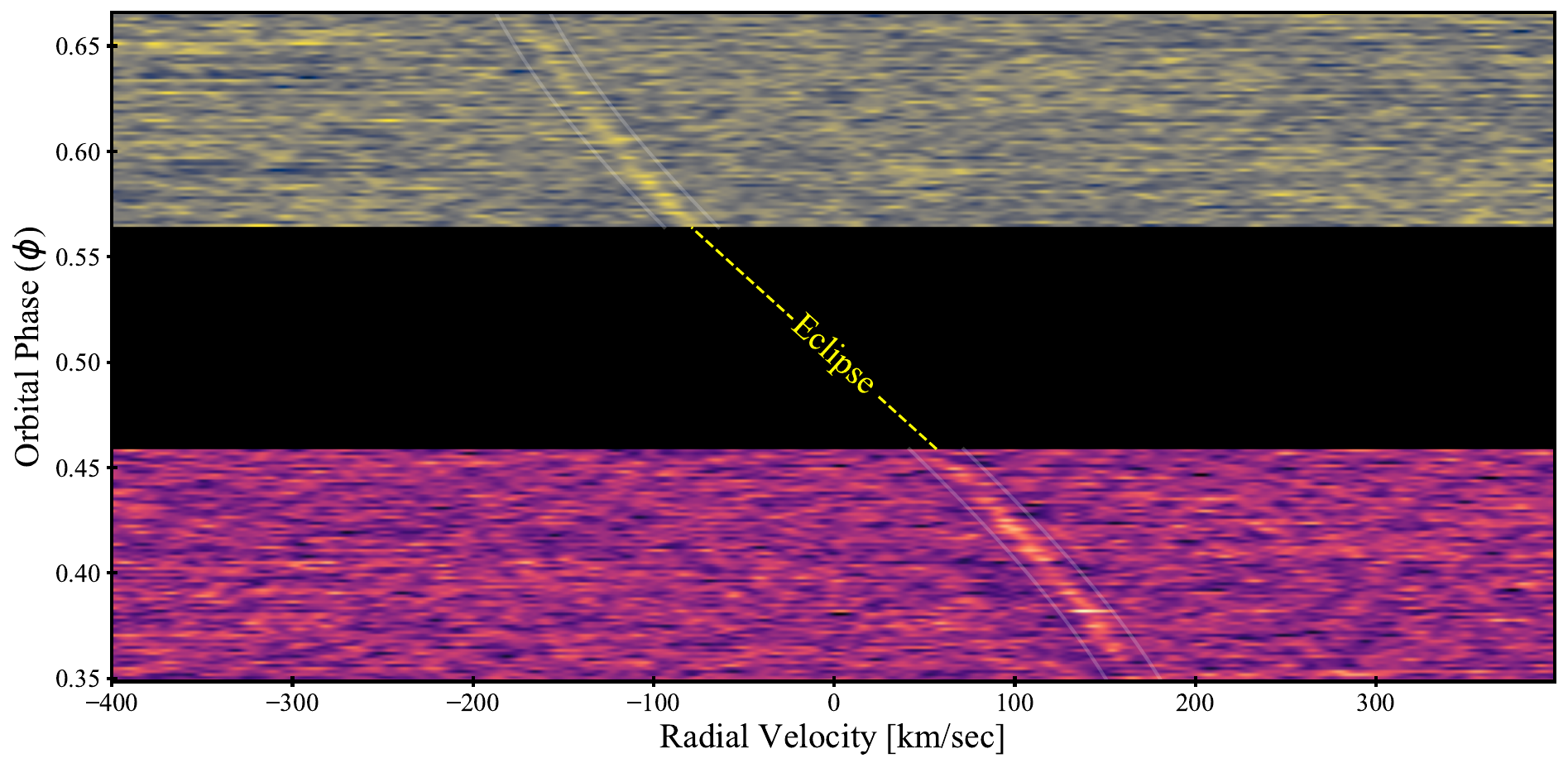}
    \caption{Trail of positive cross-correlation values following the orbital motion (along the faint white lines) of MASCARA-1\,b in the observer's rest frame. The black region denotes the phase range where the planet is eclipsed behind the host star. This trail is generated by cross-correlating a solar composition ``full'' model template (described in Section \ref{sec:cc_and_detections}) with the post-SVD data of pre- and post-eclipse geometries.}
    \label{fig:trail}
\end{figure*}

\section{Observations and data processing} \label{sec:observations}

We observed MASCARA-1\,b on two separate nights at pre- and post-secondary eclipse orbital phases using the IGRINS \citep[][]{park-igrins-cite, igrinsplp_mace2018} and IGRINS-2 \citep[][]{Oh24_IG2perfomance, Choi25_W33b} spectrographs. The pre-eclipse observations were obtained as part of the \textit{Roasting Marshmallows: Disentangling Composition \& Climate in Hot Jupiter Atmospheres through High-Resolution Thermal Emission Cross-Correlation Spectroscopy} survey program (GS-2023B-LP-206; PI: M.~Line; Figure \ref{fig:survey}) on 2021 August 05 UTC using the IGRINS spectrograph ($R \sim 45{,}000$, $1.45$--$2.45~\mu$m) mounted on the Gemini South telescope. These observations were acquired as a continuous sequence spanning approximately four hours in an AB--BA nodding pattern, with individual exposure times of 90~s. In total, we obtained 83 AB frames covering the orbital phase range $0.34 < \phi < 0.46$, where $\phi = 0.5$ corresponds to superior conjunction (Figure~\ref{fig:observations and SVD summarized}). The final six frames were taken during secondary eclipse and were therefore excluded from subsequent analysis. The median signal-to-noise ratio (SNR) per resolution element in this sequence was 190 in the H band and 195 in the K band.

Post-eclipse observations were obtained using the IGRINS-2 spectrograph ($R \sim 45{,}000$, $1.45$--$2.45~\mu$m), a near-identical successor to IGRINS, mounted on the Gemini North telescope. These data were acquired as part of the IGRINS-2 system verification program (GN-2024A-SV-103\footnote{\url{https://www.gemini.edu/instrumentation/igrins-2/igrins-2-system-verification}}; PIs: M.~Weiner-Mansfield \& E.~Deibert) on 2024 July 21. Similar to the pre-eclipse observations, these data were collected in an AB--BA nodding pattern with 90~s exposures over a continuous sequence of approximately four hours, resulting in 70 AB frames covering $0.56 < \phi < 0.66$. These observations were performed at a lower airmass (median $\approx$ 1.07) than the pre-eclipse night (median $\approx$ 1.46), therefore the post-eclipse dataset achieved a higher median SNR of 400 in the H band and 390 in the K band. A summary of the orbital phase coverage and data quality for both observing epochs is shown in Figure~\ref{fig:observations and SVD summarized}. The adopted system parameters used throughout our analysis are listed in Table~\ref{tab:parameters}.

The raw data from each night were reduced using the standard IGRINS data reduction pipeline \citep[PLP; ][]{igrinsplp_lee2016, Sawczynec25_IG2v3PLP}, which performs flat-fielding, sky subtraction, optimal spectral extraction, and wavelength calibration to produce one-dimensional spectra. Following the initial reduction, we applied a secondary wavelength refinement \citep{Line2021_W77Ab, Brogi2023, Kanumalla24_W127b, Mansfield24_W76b, Smith24_W121b, Panwar25_W122b} by performing a linear stretch and shift on each spectrum to align it with the final spectrum (taken as the reference) in the time series. This step ensures robust wavelength alignment across the sequence and is particularly important for subsequent telluric detrending. For each spectral order, 100 pixels were trimmed from both ends due to low instrumental throughput. In addition, orders severely contaminated by telluric absorption were discarded: 10 orders from the pre-eclipse dataset and 12 orders from the post-eclipse dataset. The resulting data cubes have dimensions of $44 \times 77 \times 1848$ (orders $\times$ frames $\times$ pixels) for the pre-eclipse observations and $42 \times 70 \times 1848$ for the post-eclipse observations.

The planetary signal in these data is orders of magnitude weaker than the stellar and telluric contributions. To isolate the planetary emission spectrum, we detrended each dataset using singular value decomposition (SVD) \citep{dek13, Line2021_W77Ab}. Following \cite{Line2021_W77Ab, Brogi2023, Kanumalla24_W127b, Mansfield24_W76b, Smith24_W121b}, we applied \texttt{numpy.linalg.svd} to decompose the $N_{\mathrm{frames}} \times N_{\mathrm{pixels}}$ data matrix into orthogonal components (on an order-by-order basis). The data were then separated into a low-rank reconstruction matrix (or scaling matrix), formed from the first $N_c$ singular vectors, and a residual matrix containing the remaining components. The optimal number of components retained depends on observing conditions and instrumental systematics. Based on visual inspection of residual telluric features, we adopted $N_c = 4$ for the pre-eclipse dataset and $N_c = 6$ for the post-eclipse dataset, similar to past IGRINS emission data-sets \citep[e.g.,][]{Line2021_W77Ab, Brogi2023, Smith24_W121b}. 


As a final preprocessing step, we used a telluric transmittance template from ESO SkyCalc\footnote{\url{https://www.eso.org/observing/etc/skycalc/}} and masked wavelength regions where the atmospheric transmittance was below 50\%, mitigating the residual contamination from strongly absorbed telluric regions which are imperfectly removed by SVD. An example of the detrending process for a spectral order with heavy telluric contamination is shown in the lower-right panel of Figure~\ref{fig:observations and SVD summarized}.

\section{Cross-correlation analysis: Detection of volatile and refractory gases} \label{sec:cc_and_detections}


\begin{figure*}
    \centering
    \includegraphics[width=\textwidth]{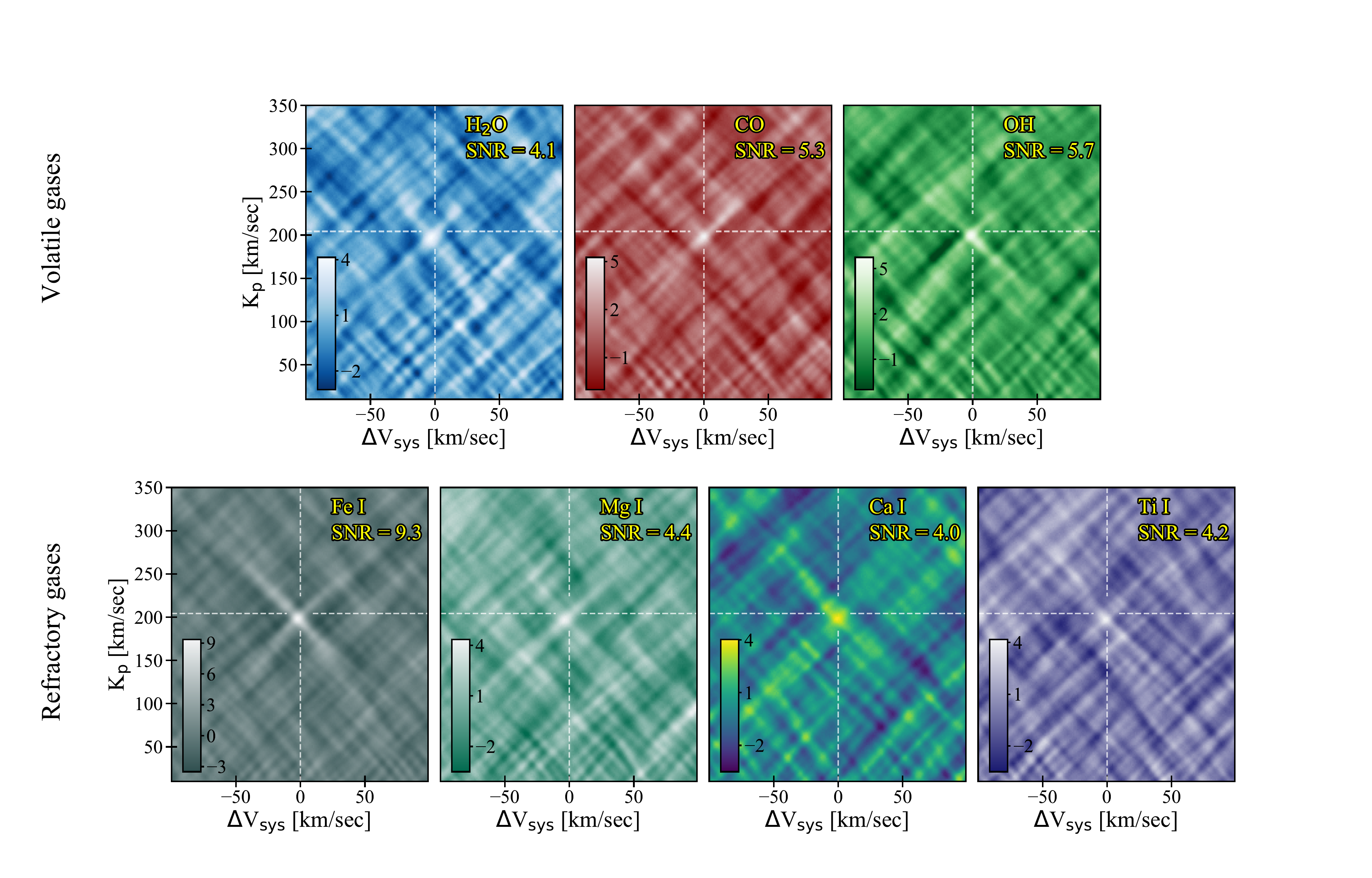}
    \caption{Detections from our cross-correlation analyses shown as K$_\mathrm{p}$-$\Delta$V$_\mathrm{sys}$ maps. Each map has been generated using a solar composition RCTE model template containing a single gas opacity along with the continuum. The top row shows the detection of volatile gases i.e., H$_2$O, CO and OH. The bottom row shows our detections of refractory gases i.e., neutral Fe, Mg, Ca, and Ti. The white perpendicular dotted lines denote the literature K$_\mathrm{p}$ and $\Delta$V$_\mathrm{sys}$ values ($\Delta$V$_\mathrm{sys}$ = 0 corresponds to the planetary rest frame). The red cross in all these maps denotes the peak cross-correlation value.
    }
    \label{fig:CCmaps}
\end{figure*}

We utilize cross-correlation to identify the presence of atmospheric species in the residual pre- and post-eclipse datasets \citep[e.g.,][]{sne10,Bir18, BL19}. We first describe the synthetic model templates, followed by the cross-correlation procedure used to detect individual atomic and molecular species.

We generated model templates for the initial cross-correlation analysis using the \texttt{ScCHIMERA} 1D-radiative-convective-thermochemical equilibrium (1D-RCTE) solver \citep[e.g.,][]{Piskorz18_KELT2Ab, Mansfield2018, Wiser25_W80b, Welbanks24_W107b}. Given the system parameters listed in Table~\ref{tab:parameters}, along with assumptions about atmospheric composition, incident stellar flux, and internal temperature ($T_{\mathrm{int}}$), \texttt{ScCHIMERA} computes a disk-averaged dayside temperature--pressure (T--P) profile and the corresponding volume mixing ratios (VMRs) under the RCTE assumption. For our initial model set used for cross-correlation analysis, we assumed a solar atmospheric composition ([M/H]$=0$, C/O$=0.55$), a $T_{\mathrm{int}} =$  650 K following \cite{thorngren-fortney-2019}, and adopted a heat redistribution factor of $f = 2.2$ following \citet{Parmentier2021}. A \texttt{PHOENIX} \citep{Husser13_Phoenix} stellar model (at T$_\mathrm{eff}$ $\approx$ 7490 K and log$_{10} g_\star$ $\approx$ 4.09 c.g.s) was used as the incident stellar spectrum.

The resulting T--P and gas VMR profiles were then processed using the GPU--enabled \texttt{CHIMERA} radiative transfer module \citep{Line2021_W77Ab, bell_wasp80b_ch4, Wiser25_1Dvs3D}  to generate high-resolution ($R = \lambda/\Delta\lambda = 250{,}000$) emission spectra of the planetary atmosphere. These models included line opacity from a suite of volatile and refractory species dominant at MASCARA-1~b
temperature regime, including H$_2$O, CO, OH, Fe\,\textsc{i}, Mg\,\textsc{i}, Ca\,\textsc{i}, Si\,\textsc{i}, Ti\,\textsc{i}, V\,\textsc{i}, Cr\,\textsc{i}, TiO, and VO. Continuum opacity sources from collision-induced absorption (H$_2$--H$_2$, H$_2$--He), bound-free absorption (H$^-$), and free-free absorption were also included. The references for these opacity sources are given in Appendix Table \ref{tab:opacity-references}. To account for the solid-body rotation of MASCARA-1\,b ($v_{\mathrm{rot}} = 3.55~\mathrm{km~s^{-1}}$; Table \ref{tab:parameters}), the planetary spectrum was convolved with a rotational broadening kernel, followed by convolution with a Gaussian instrumental profile matched to the resolving power of IGRINS and IGRINS-2 (assumed to be a constant R=45,000 across each order).

An analogous procedure was applied to the stellar spectrum, including convolution with a stellar rotational kernel ($v\sin i_\star = 101.7~\mathrm{km~s^{-1}}$; Table \ref{tab:parameters}) and the appropriate instrumental profile. We also applied a low pass gaussian filter (using \texttt{scipy.ndimage.gaussian\_filter1d}) with a width of 1000 pixels in each order to the convolved stellar spectrum, smoothing out remaining sharp stellar features. The cross-correlation templates were constructed as planet-to-star flux ratio spectra, $(F_p/F_\star)$, scaled by the squared ratio of planetary to stellar radii, $(R_p/R_\star)^2$. To isolate the contribution from individual opacity sources, we additionally generated single gas species templates (with continuum opacity) by including only one molecular or atomic absorber at a time.

The SVD detrending applied during preprocessing can modify the intrinsic shape and amplitude of the planetary signal. To ensure a consistent comparison between data and models, we injected the synthetic model templates into the scaling matrix and processed them using the same detrending procedure applied to the data. This approach yields filtered model templates that have undergone the same linear transformations as the planetary signal in the observations, enabling unbiased cross-correlation detections \citep[e.g.,][]{BL19, Gibson22_W121b}.

The planet's signal in our observations is expected to follow a circular Keplerian orbit \citep[][Equation 4]{Bir18} parameterized by semi-amplitude of radial velocity curve (K$_\mathrm{p}$), systemic velocity of MASCARA-1 system ($V_{\mathrm{sys}}$), and barycentric velocity of the observer ($V_{\mathrm{bary}}$). Therefore, in each order, the post-SVD residual spectrum was cross-correlated with Doppler-shifted model templates evaluated at the corresponding line of sight velocity $V_p(\phi)$ at given phase ($\phi$), producing a time-dependent trail of cross-correlation values following the expected orbital motion in the observer's rest frame in Figure \ref{fig:trail}.

To determine the individual gas detection signal-to-noise ratios (SNR), we computed two-dimensional K$_\mathrm{p}$--$\Delta V_{\mathrm{sys}}$ maps by evaluating the cross-correlation over a grid of K$_\mathrm{p}$ and $\Delta V_{\mathrm{sys}}$ values and summing coherently across all spectral orders and frames. The resulting grid of cross-correlation values were sigma clipped at 3$\sigma$ level to remove any outliers and then normalized by the standard deviation of the entire grid to produce a map of SNR values. This procedure collapses the orbital trail into a localized peak in velocity space when the model parameters match the true planetary motion, a standard diagnostic in this type of analysis \citep[e.g.,][]{Line2021_W77Ab, Brogi2023, Kanumalla24_W127b}.

Figure~\ref{fig:CCmaps} presents the resulting K$_\mathrm{p}$--$\Delta V_{\mathrm{sys}}$ maps for each species we have investigated. We detect robust signals from the volatile gases H$_2$O (SNR$=$4.1), CO (5.3), and OH (5.7), as well as from the refractory species Fe\,\textsc{i} (9.3), Mg\,\textsc{i} (4.4), Ca\,\textsc{i} (4.0), and Ti\,\textsc{i} (4.2). It is worth noting, that while several detections are below an SNR=5, all maps in Figure \ref{fig:CCmaps} have their dominant peak value near the nominal K$_\mathrm{p}$--$\Delta V_{\mathrm{sys}}$. Most of the noise that reduces the overall detection SNR is due to the cross-hatching arising from aliasing of the planetary lines in the cross-correlation method \citep[e.g.,][]{Borsato2023_KELT9b}.  

Among the detected volatiles, H$_2$O exhibits a comparatively weaker signal, consistent with partial thermal dissociation at the high dayside temperatures of MASCARA-1\,b \citep[e.g.,][]{Parmentier2018_WASP121b_thermal_dissociation}; the simultaneous detection of OH supports this interpretation. Similar trends have been observed in previous high-resolution studies of UHJs  \citep[e.g.,][]{Nugruho2021_WASP33b_OH, Brogi2023, Smith24_W121b, Sanchez26_W189b, Bazinet25_W121b_NIRPS}. Among the refractory species, Fe\,\textsc{i} yields the strongest detection, reflecting the large number of Fe\,\textsc{i} lines present within the IGRINS wavelength coverage. We do not detect statistically significant signals from Si\,\textsc{i} or V\,\textsc{i} (Appendix Figure \ref{fig:nondetections}). Also,  despite a previous detection by \citet{Scandariato23_M1b}, we do not recover Cr \textsc{i}. This is likely because the strongest Cr \textsc{i} lines are located at optical wavelengths, whereas our H-band data suffers from low signal-to-noise ratios and heavy telluric masking in the relevant orders.

\section{Retrieval analysis: composition, thermal profile, and velocity offsets} \label{sec:retrievals}

\begin{figure*}[htbp]
    \centering
    \includegraphics[width=\textwidth]{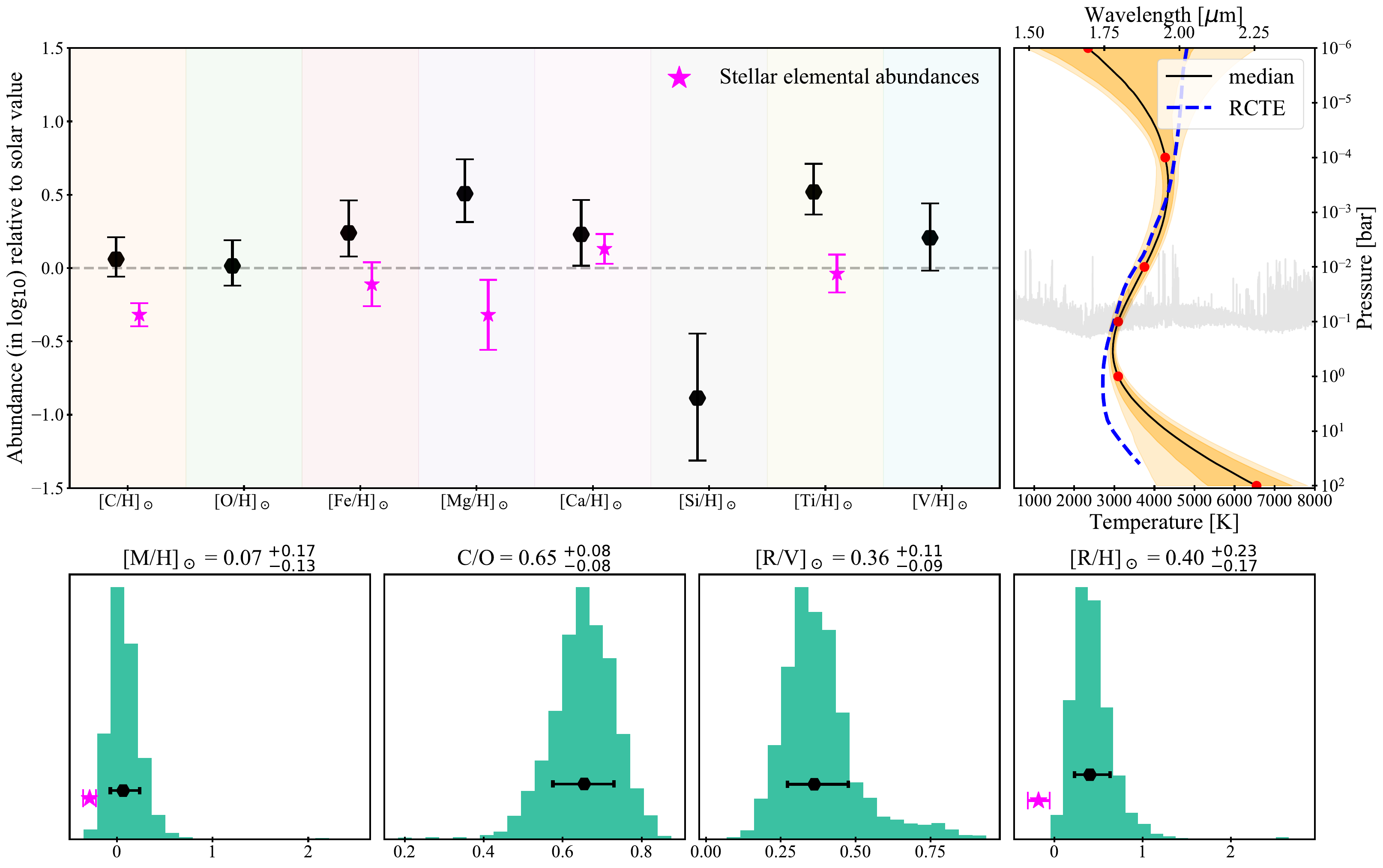}
    \caption{
    Summary of the fiducial chemical equilibrium retrieval results. Our obtained medians and 1$\sigma$ errors on the elemental abundances relative to solar are shown in the top left panel. Elemental abundances are expressed as logarithmic multiplicative factors relative to solar composition. The solar abundance is shown for reference as the dashed gray horizontal line. For elements with measured stellar abundances, we have shown the stellar values in star shaped markers with magenta color. The top right panel shows the retrieved median T--P profile (solid black) with 1$\sigma$ ($\sim$68\%) and 2$\sigma$ ($\sim$95\%) confidence regions shaded in lighter orange; red points mark the pressure nodes used in the Bézier spline interpolation (see text). The gray background spectrum indicates the wavelength-dependent photospheric pressure levels (defined at $\tau = 2/3$), and the blue dashed line shows the radiative–convective–thermochemical equilibrium (RCTE) profile generated with \texttt{ScCHIMERA} for best fit composition. The bottom-row panels display derived bulk atmospheric properties computed from the retrieved elemental abundances. We obtain a solar atmospheric metallicity ([M/H]$_\odot$), a solar C/O ratio, a moderately super-solar refractory-to-volatile ratio ([$\mathcal{R}/\mathcal{V}$]$_\odot$), a super-solar refractory abundance([$\mathcal{R}$/H]$_\odot$). In these panels, we also show calculated stellar estimates in star shaped markers for reference. We note that since we do not obtain statistically significant CC signals from Si \textsc{I} and V \textsc{I}, elemental abundances of Si and V must be interpreted with caution. We have excluded these abundances in the derived abundance ratios above. }
    \label{fig:retrieval-summarized}
\end{figure*}

\subsection{Atmospheric elemental abundances and temperature--pressure profile}

To obtain quantitative constraints on the atmospheric composition (parameterized by elemental abundances) and thermal structure,  we apply an atmospheric retrieval analysis using the \cite{BL19} likelihood formalism. We couple the chemical equilibrium solver \texttt{FASTCHEM} \citep[][]{Stock2018_fastchem, kitzmann2024} with a flexible temperature–pressure parameterization and \texttt{CHIMERA} forward model, and sample the parameter space with \texttt{pymultinest} \citep[][]{Feroz2009, Buchner2014}. Within this chemically consistent framework, we fit directly for the logarithmic elemental abundances relative to solar values ([X/H], where X is C, O, Fe, Mg, Ca, Si, Ti, and V). We used the Bézier-spline T--P parameterization described in \citet{Smith24_W121b, Panwar25_W122b} with six fixed nodes at pressures uniformly spaced between \(10^{2}\) and \(10^{-6}\,\mathrm{bar}\). This parameterization allows the flexibility to capture vertical temperature gradients with a relatively small number of free parameters. We also fit the HRCCS-specific velocity parameters as  $\Delta$\(K_{p}\) and $\Delta$\(V_{\text{sys}}\) for each night separately. We adopted uniform priors of all the retrieval parameters. The prior bounds on the logarithmic elemental abundances (relative to solar value) are between -3 to 3, on the six temperature nodes are between 500 to 8000 K, and on the two velocity parameters is -50 to 50 km/sec. In the forward model used in the retrieval, we include all the gas opacities from Section \ref{sec:cc_and_detections} except for TiO and VO to avoid any biases due to inaccurate line lists at high-resolution \citep[][]{Hoeijmakers15_HD209_TiO, deRegt22_VO}. We used 300 live points, a log-evidence tolerance of 0.5, and a sampling efficiency of 0.8 with \texttt{pymultinest} to sample the posterior distribution. The radiative transfer calculations were accelerated on an NVIDIA A100 GPU, while likelihood evaluations were parallelized over 12 Intel Broadwell CPUs. The retrieval required approximately five weeks to complete, involving $\approx$ 1.1 million likelihood evaluations.

Figure~\ref{fig:retrieval-summarized} summarizes the nominal retrieved constraints, with the full posterior distributions shown in Appendix Figure~\ref{fig:appendix-full-corner}. From the posterior sample, we reconstruct the median T--P profile by sampling 1000 random draws and computing the corresponding temperature structure for each sample. Following this, the median T--P profile and 1$\sigma$ error (at each pressure layer) were obtained. These are shown in the right inset of Figure~\ref{fig:retrieval-summarized}. 

We retrieve solar abundances for C ([C/H]$= 0.06^{+0.15}_{-0.12}$), O ([O/H]$= 0.01^{+0.17}_{-0.13}$), and V ([V/H]$= 0.21^{+0.23}_{-0.22}$), significantly super-solar abundances for Mg ([Mg/H]$= 0.51^{+0.23}_{-0.19}$; $\approx$3.9 $\times$ solar) and Ti ([Ti/H]$= 0.52^{+0.19}_{-0.15}$; $\approx$3.3 $\times$ solar), a moderately super-solar abundance for Fe ([Fe/H]$= 0.24^{+0.22}_{-0.16}$; $\approx$1.5~$\times$ solar) and Ca ([Ca/H]$= 0.23^{+0.23}_{-0.22}$; 2.1 $\times$ solar), and a significantly sub-solar abundance for Si ([Si/H]$= -0.89^{+0.44}_{-0.43}$; $ \approx$0.2~$\times$ solar). The typical precision on individual elemental abundances is $\sim$0.2 dex, with the exception of Si, which remains less well constrained.

It is important to highlight that we do not detect any confident signals from Si\,\textsc{i} and V\,\textsc{i} at the nominal velocities in the $\Delta$K$_\mathrm{p}$-$\Delta$V$_\mathrm{sys}$ maps. However, both maps show a weak signal (SNR $<$ 3) near the literature K$_\mathrm{p}$ and $\Delta$V$_\mathrm{sys}$ values (Appendix Figure~\ref{fig:nondetections}). Therefore, we interpret atmospheric Si and V elemental abundances with caution. The obtained constraints on Si and V elemental abundances are likely due to inclusion/removal of the opacity of Si\,\textsc{i} and V\,\textsc{i} affecting the overall template spectrum as discussed in \cite{Smith24_W121b}.  However, more data (preferably in the optical wavelengths) will prove highly essential to confirm these tentative signals and their abundances.

Using the available stellar elemental abundances reported by \citet{Saffe2021} (Table \ref{tab:parameters}), we compare the retrieved elemental abundances relative to their stellar values. We find that the atmospheric abundances of C ([C/H]$_\star$ = $0.38^{+0.17}_{-0.14}$), Mg ([Mg/H]$_\star$ = $0.85^{+0.33}_{-0.31}$), and Ti ([Ti/H]$_\star$ = $0.55^{+0.23}_{-0.20}$) are super-stellar. Ca ([Ca/H]$_\star$ = $0.10^{+0.25}_{-0.24}$) is consistent with the stellar value. Fe ([Fe/H]$_\star$ = $0.36^{+0.25}_{-0.22}$) is only moderately enhanced relative to the host star.  

We compute the atmospheric metallicity ([M/H]) by converting each retrieved elemental abundance from logarithmic to linear units relative to hydrogen and summing their contributions. The C/O ratio is derived analogously from the total atmospheric carbon and oxygen abundances. From this procedure, we obtain a solar atmospheric metallicity i.e., [M/H]$_\odot$ $= 0.07^{+0.17}_{-0.13}$, C/O $= 0.65^{+0.08}_{-0.08}$ that is consistent with solar value i.e., 0.59~\citep[][]{Asplund2021_solar}. Notably, these estimates are consistent with \cite{Ramkumar2025} analysis of MASCARA-1\,b using CRIRES+ observations (Appendix Figure~\ref{fig:ratios-compared}). Excluding Si and V abundances, we also obtain an enriched total refractory abundance ([$\mathcal{R}$/H]$_\odot$ = $0.40^{+0.23}_{-0.17}$; $\approx$2.5 $\times$ solar value and $\approx$3.8 $\times$ stellar value) and a refractory-to-volatile enhancement of [$\mathcal{R}/\mathcal{V}$]$_\odot =$ $0.36^{+0.11}_{-0.09}$ ($\approx$2.3 $\times$ solar value) in MASCARA-1\,b's atmosphere. Together, these quantities provide multi-dimensional compositional constraints that enable our discussion of MASCARA-1b’s formation history in Section~\ref{sec:formation}.

\subsection{Investigating species-dependent dynamics in MASCARA-1\,b's atmosphere}

Thermochemical and dynamical inhomogeneities across the dayside hemisphere can cause the observed atmospheric signal to deviate from the simple Keplerian expectation. In particular, longitudinal temperature variations and atmospheric winds can introduce measurable offsets in the retrieved $K_{\mathrm{p}}$ and $V_{\mathrm{sys}}$ \citep[e.g.,][]{Cont21_W33b,Brogi2023,Smith24_W121b,Seidel25_W121b,nortmann-w127b-crires+}. These offsets can encode information about atmospheric circulation and spatially varying emission, but are also influenced by errors on assumed orbital solution \citep[i.e., ephemeris and eccentricity][]{peter_wasp77Ab_igrins, Pino2022_KELT9b}. We adopted the latest ephemeris and orbital parameters from \cite{Hooton2022} to avoid any biases from incorrect orbital solution. 

To investigate potential species-dependent dynamical signatures in MASCARA-1\,b, we performed additional retrievals using the single species best-fit model templates from our fiducial analysis and allowed $\Delta $K$_{\mathrm{p}}$, $\Delta $V$_{\mathrm{sys}}$, and $\log a$ to vary. Here, $\Delta $K$_{\mathrm{p}}$ and $\Delta $V$_{\mathrm{sys}}$ capture deviations from the nominal orbital solution, while $\log a$ scales the line contrast and accounts for geometric dilution of the dayside emission. In general, $\log a$ accommodates longitudinal brightness inhomogeneities in disk-integrated spectra. However, since we use best fit models templates in this exercise, we retrieved log $a$ $\approx$ 0 as the best fit model is already optimized to fit the line strengths.

\begin{figure}[ht!]
    \centering
    \includegraphics[width=\columnwidth]{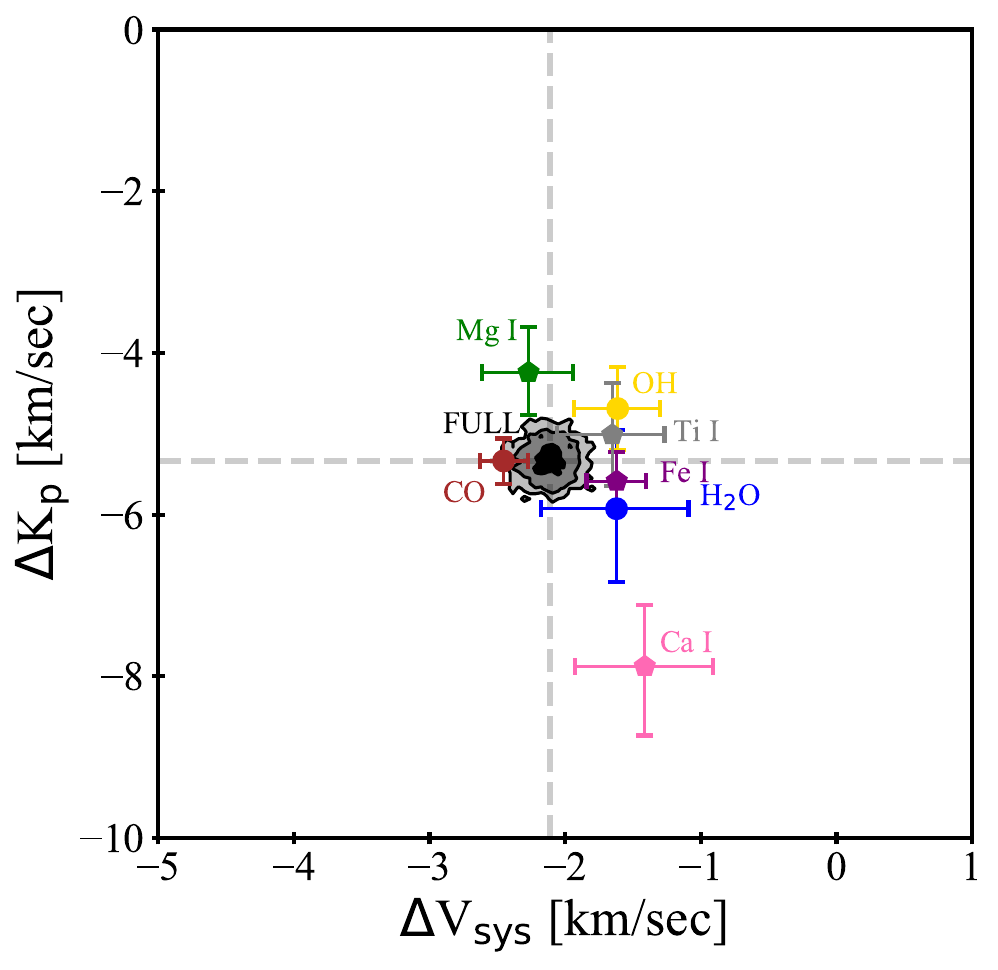}
    \caption{Distribution of velocity offsets ($\Delta K_\mathrm{p}$ and $\Delta V_\mathrm{sys}$) retrieved using best-fit single species templates assuming the fiducial composition. The dashed lines indicate the $K_\mathrm{p}$ and $V_\mathrm{sys}$ offset of the full model template. Negative offsets in $K_\mathrm{p}$ are expected in eclipse geometry \citep[e.g.,][]{Wardenier25_3D} and, in this case, can be explained by MASCARA-1b’s solid body rotation combined with an equatorial jet velocity of $\approx 2~\mathrm{km\,s^{-1}}$. Along the $\Delta V_\mathrm{sys}$ direction, we find weak evidence for non-zero offsets. This could potentially imply sluggish atmospheric winds potentially due to a strong drag mechanism \citep[e.g.,][]{Rauscher_Menou_2012_drag, Beltz2022_drag_WASP76b, Bell2021} but the uncertainty in the systemic velocity prevents us from drawing a firm conclusion.} 
    \label{fig:velocity_offsets}
\end{figure}

Figure \ref{fig:velocity_offsets} shows the distribution of $\Delta$K$_\mathrm{p}$ and $\Delta$V$_\mathrm{sys}$ values obtained from above-mentioned retrievals using single model templates. As investigated in \cite{Wardenier25_3D}, deviations from expected K$_\mathrm{p}$ occur due to the planet's rotation, and the upper limit on $\Delta$K$_\mathrm{p}$ can be derived to be $\lvert$$\Delta$K$_\mathrm{p}$$\rvert$ $\leq$ $\varv_\mathrm{eq} + \varv_\mathrm{jet}$. Here, $\varv_\mathrm{eq}$ denotes the equatorial solid body rotation velocity and $\varv_\mathrm{jet}$ denotes the equatorial jet speed. Given the small propagated uncertainty on K$_\mathrm{p}$ (Table \ref{tab:parameters}), in our case, this means that on top of solid body rotation ($\approx$ 3.5 $~\mathrm{km~s^{-1}}$), a modest jet speed of $\approx$ 2 $~\mathrm{km~s^{-1}}$ can explain $\Delta$K$_\mathrm{p}$ for all species except Ca\,\textsc{i}. These jet speeds are indeed predicted in Global Circulation models (GCMs) of UHJs when atmospheric drag is included \citep[e.g.,][]{Tan_Komacek_2019_GCM, Komacek_Showman_2020_GCMs, Beltz2022_drag_WASP76b, Roth2024_GCM_grid}.  The significant K$_\mathrm{p}$ offset of Ca\,\textsc{i} warrants further investigation from the optical wavelengths where Ca\,\textsc{i} and Ca\,\textsc{ii} have stronger spectral footprint than H $\&$ K bands.

Due to the high uncertainty in the reported systemic velocity \citep{Talens2017, casasayas2019}, we use the relative offset in $\Delta$V$_\mathrm{sys}$ from the full model template to interpret species dependent dynamics in MASCARA-1b's atmosphere. In Figure \ref{fig:velocity_offsets}, we have shown 1$\sigma$ error bars for visual clarity and it seemingly shows a noticeable V$_\mathrm{sys}$ offset for certain gases (for e.g., CO, Fe\,\textsc{i}, and Ca\,\textsc{i}). However, within a 2 $\sigma$ region, all the contours overlap making it challenging to draw any meaningful conclusions from the $\Delta$ V$_\mathrm{sys}$ offsets. Therefore, although we detected an average offset in V$_\mathrm{sys}$ among all the gases, the variations among the gases themselves are not particularly significant. It is perhaps interesting to note that \cite{Bell2021} have reported a hotspot offset of 6$\pm$11$^\circ$W. If the absence of a significant hotspot offset is a result of strong drag mechanism \citep[][]{Rauscher_Menou_2012_drag, Beltz2022_drag_WASP76b}, the sluggish winds implied by our V$_\mathrm{sys}$ offsets agree with such reasoning.

\section{Link to planet formation enabled by the refractory and volatile abundances}
\label{sec:formation}

\begin{figure}[htbp]
    \centering
    \includegraphics[width=\columnwidth]{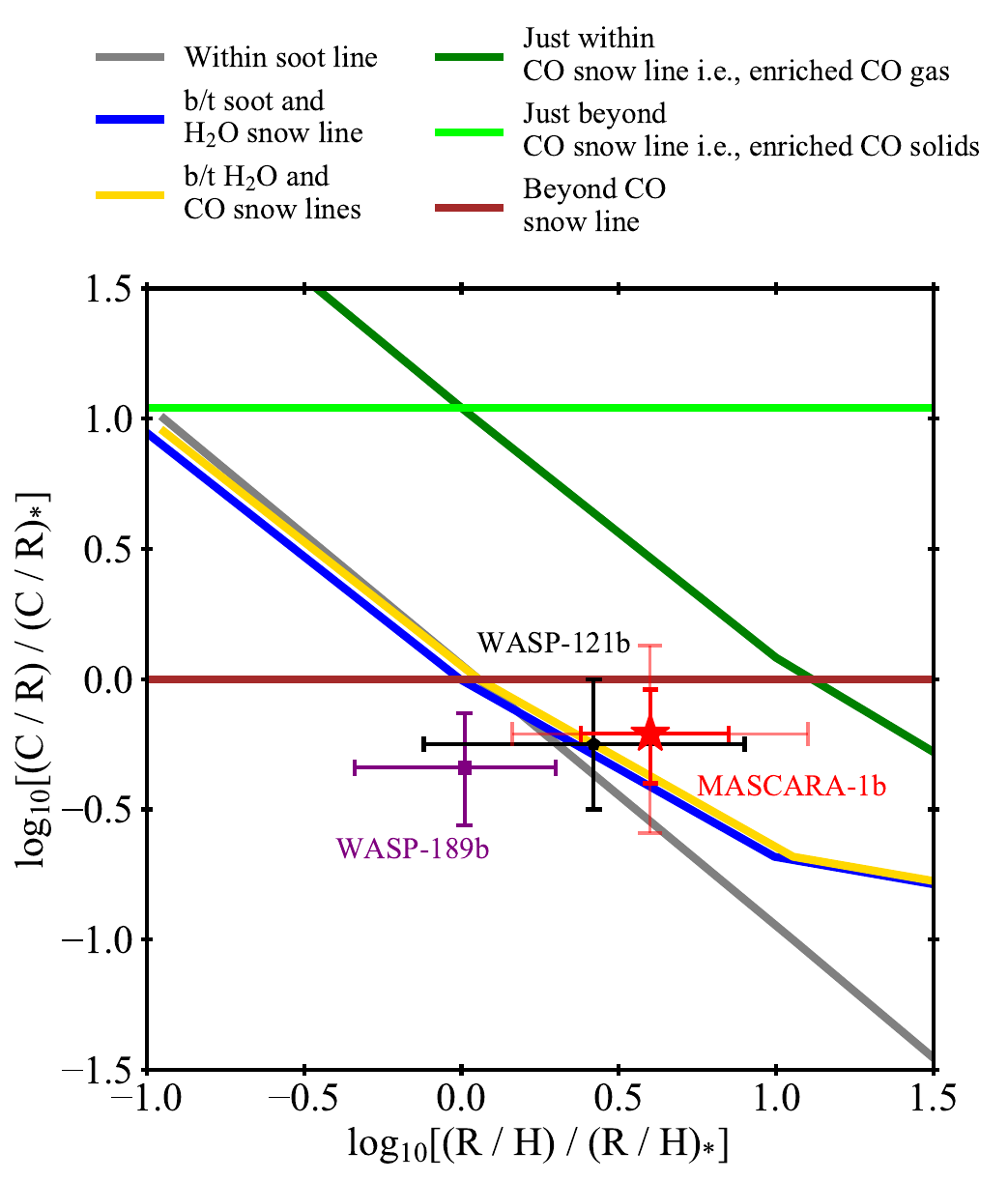}
    \caption{Observed elemental ratios of MASCARA-1b (red), WASP-121\,b (black), and WASP-189\,b (purple) compared with the tracks of formation scenarios from \cite{Chachan2023}. Colored lines indicate predicted C/$\mathcal{R}$ vs $\mathcal{R}$/H trends for planets forming at different disk locations relative to major snow lines. Within the 1$\sigma$ region (opaque error bar), the scenario that most agrees with our results is that MASCARA-1b formed either between the soot and H$_2$O snowline or between H$_2$O and CO snowlines. Within a 2$\sigma$ region (error bar with less opacity), our results are also consistent with a formation scenario occurring within the soot line and beyond the CO snowline.}
    \label{fig:formation}
\end{figure}

The key additional constraint provided by this work is the refractory-to-volatile ratio ($\mathcal{R}/\mathcal{V}$), which adds an independent compositional axis beyond bulk metallicity and C/O for interpreting formation location. The inclusion of refractory abundances enables a more discriminating comparison with formation models. We therefore compute elemental ratios between refractory and volatile species and compare our results within the framework of \citet{Chachan2023}. Under the assumption of a fully mixed envelope, these models predict the abundance ratios C/$\mathcal{R}$, O/$\mathcal{R}$, and $\mathcal{R}$/H for various accretion locations relative to major snow lines. To estimate $\mathcal{R}$/H and C/$\mathcal{R}$ for MASCARA-1\,b, we use the retrieved abundances of refractory elements in the context of the available stellar abundance measurements (Fe, Mg, Ca, Ti).

Figure \ref{fig:formation} compares our measured abundance ratios with the formation tracks of \citet{Chachan2023}. Within 1$\sigma$ confidence, the retrieved $\mathcal{R}$/H and C/$\mathcal{R}$ ratios are most consistent with accretion occurring either between the soot and H$_2$O snowlines or between the H$_2$O and CO snowlines.  At the 2$\sigma$ level, formation scenarios interior to the soot line or exterior to the CO snowline remain statistically plausible. However, it is interesting to point out that the sub-stellar C/$\mathcal{R}$, super-stellar C/H, and super-stellar $\mathcal{R}$/H imply that MASCARA-1~b had significant accretion of solids but likely did not accrete C-depleted solids which exist inside the soot line. In principle, the O/R ratio provides an additional diagnostic for distinguishing between these formation pathways in Figure~\ref{fig:formation} \citep[e.g.,][]{Pelletier2025_wasp121b, Smith24_W121b}. However, the lack of any stellar O/H measurement for MASCARA-1 limits us from overcoming the degeneracies between the possible formation scenarios.

Among all the possible formation scenarios, if MASCARA-1\,b seemingly formed interior to the H$_2$O snowline, we depart from the conventional expectation that giant planets predominantly form further out in the disk (beyond H$_2$O snowline), where solid surface densities are sufficient to form their $\sim10$–15 $M_{\oplus}$ cores. However, alternative pathways have been proposed. For example, \citet{Bailey_Batygin_2018_insitu_formation} and \citet{Batygin2016_insitu} demonstrate that giant planets may originate in inner disk regions as super-Earth–mass cores that undergo runaway gas accretion prior to disk dispersal. Conversely, \cite{Hand_Helled_2021_migration} discuss that formation beyond the H$_2$O snowline followed by inward disk-driven migration can lead to enhanced accretion of refractory-rich solids interior to the snowline, thereby enriching the refractory abundance as we have obtained in this study. Both scenarios are consistent with radial velocity surveys and theoretical frameworks indicating that giant planets are common between 1–10 au \citep{Fulton2021,Chachan2021}.

Within a 2$\sigma$ level, accretion beyond the CO snowline followed by inward migration remains formally consistent with the retrieved abundance ratios. Such a pathway could also accommodate the observed spin–orbit misalignment of MASCARA-1\,b, as high-eccentricity migration mechanisms (such as Kozai–Lidov cycles or secular interactions with additional companions) can lead to both orbital shrinkage and obliquity excitation \citep[e.g.,][]{Kozai_1962, Lidov_1962, Fabrycky_Tremaine_2007_orbital_shrinkage, Wu_2007_secular_interactions}. However, the chemical constraints alone do not uniquely distinguish between disk-driven migration and high-eccentricity migration scenarios. Moreover, it is highly important to note that MASCARA-1 lies above the Kraft break \citep[e.g.,][]{Kraft1967, Wang26_KraftBreak}. Therefore, a definitive investigation of MASCARA-1~b's dynamical history requires independent constraints beyond our composition analysis.

\section{An emerging sample of Ultra-Hot Jupiters with refractory-to-volatile ratios}

\label{sec:UHJ_sample}

Our interpretation of MASCARA-1\,b’s formation is broadly consistent with previous analyses of WASP-121\,b \citep[from ][]{Smith24_W121b} and WASP-189\,b \citep[from ][]{Sanchez26_W189b}, both spin–orbit misaligned UHJs and key targets of the Roasting Marshmallows survey. From \cite{Smith24_W121b}, the formation of WASP-121\,b is best explained by accretion between the soot and H$_2$O snowlines. We also note that \cite{Pelletier2025_wasp121b} arrive at an opposite conclusion that WASP-121\,b is volatile enriched and therefore formed beyond the CO snowline. However, subsequent JWST/NIRISS analysis of \cite{Pelletier2026_WASP121b_JWST} is consistent with the conclusions of \cite{Smith24_W121b} (when geometric albedo is fixed at 0) and \cite{Pelletier2025_wasp121b} (when geometric albedo is freely fitted). We will compare our results to \cite{Smith24_W121b} owing to the consistency of our observations, analysis pipelines, and modelling frameworks. For WASP-189\,b, its formation is most consistent with accretion between the H$_2$O and CO snowlines, though it also permits formation interior to the H$_2$O snowline within uncertainties. The allowed formation scenarios for MASCARA-1\,b overlap with these results, raising an interesting question if these systems potentially could share a common formation pathway.

\begin{figure*}[htbp]
    \centering
    \includegraphics[width=\textwidth]{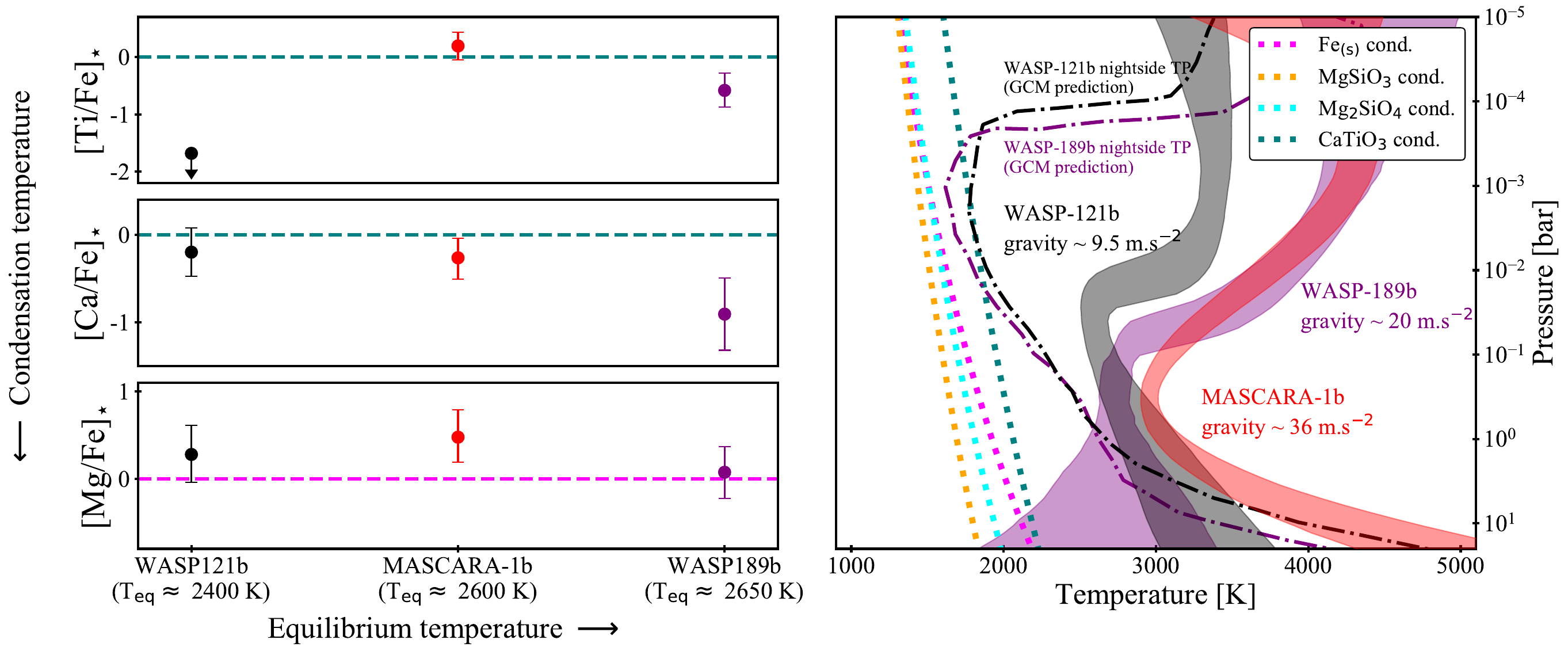}
    \caption{Refractory elemental ratios in MASCARA-1\,b along with WASP-121\,b \citep[from][]{Smith24_W121b} and WASP-189\,b \citep[from][]{Sanchez26_W189b}. We find stellar values for Ti/Fe and Ca/Fe. However, the Mg/Fe is super-stellar but consistent with stellar value at 2$\sigma$ confidence. From these, we conclude that there is no strong evidence for nightside cold trapping. On the right panel, we have compared the retrieved thermal structures for WASP-121~b, WASP-189~b, and MASCARA-1~b. Based on the predictions of nightside thermal profiles  of WASP-189\,b and WASP-121\,b \citep{Lee2022_GCM_WASP121b_WASP189b}, nightside condensation can explain their stellar Mg/Fe and depletion of refractory ratios (Ca/Fe, Ti/Fe in WASP-189b and Ti/Fe in WASP-121~b). Even though MASCARA-1~b is as hot as WASP-189~b and has higher gravity, it does not show evidence for nightside cold trapping. However, sophisticated GCMs will prove essential to further explain MASCARA-1\,b's refractory ratios.}
    \label{fig:fetomg}
\end{figure*}

Figure \ref{fig:formation} also suggests a tentative preference for refractory elements being enriched over volatiles for the three spin–orbit misaligned UHJs, MASCARA-1\,b and WASP-121\,b being clear examples. In contrast, KELT-20\,b, an aligned UHJ, has been reported to exhibit a volatile-rich composition and a sub-solar $\mathcal{R}/\mathcal{V}$ ratio by \cite{Chachan25_KELT20b} from HST/WFC3-UVIS observations. This qualitative contrast raises the possibility that refractory enrichment may correlate with orbital misalignment, although the current sample size is too small to establish a statistically robust trend. Additional systems provide a more nuanced picture. For example, \cite{Lothringer25_W178b} use HST/WFC3 and JWST/NIRSpec observations to report mixed indications for WASP-178\,b, another misaligned UHJ, depending on the specific abundance tracer employed (e.g., Si/O versus Si/C). These results highlight the importance of homogeneous analyses across a broader sample of UHJs to determine whether refractory enrichment and spin–orbit misalignment are physically connected or coincidental.

At present, the three Roasting Marshmallows targets with $\mathcal{R}$/$\mathcal{V}$ constraints  occupy a relatively narrow region of $\mathcal{R}$/H–C/$\mathcal{R}$ parameter space (Figure \ref{fig:formation}). Expanding this analysis to additional UHJs, particularly systems with well-constrained stellar abundances and diverse spin–orbit architectures, will be essential for testing whether refractory enrichment is a common feature of misaligned UHJs or merely an artifact of small-number statistics. Such pursuits will lay a necessary foundational work that can be continued in the upcoming era of Extremely large telescopes (ELTs) to enable a  population level inference on UHJ formation.

\section{Refractory elemental ratios show no strong evidence for nightside cold~trapping}
\label{sec:coldtrapping}

Several theoretical works suggest that refractory element ratios (e.g., Fe/Si, Mg/Si) in giant planet atmospheres reflect stellar values \citep[e.g.,][]{Thiabaud_2015_Elemental_ratios, Dorn2015_rocky_interior_structures, adibekyan2024}, and recent observational studies have supported this prediction \citep[e.g.,][]{Gibson22_W121b, Bazinet25_W121b_NIRPS, Sanchez26_W189b}.  We can therefore generally assume that significant deviations from stellar ratios likely arise from atmospheric processes (as opposed to planet formation processes) such as night-side cold trapping \citep[e.g.,][]{Parmentier2013_coldtrap_HD209}. 
To test this hypothesis, we compare atmospheric Ti/Fe, Ca/Fe, and Mg/Fe relative to their stellar values. For comparison, we performed the same analysis using the published IGRINS derived values for WASP-189\,b \citep{Sanchez26_W189b} and WASP-121\,b \citep{Smith24_W121b}.  

In WASP-121\,b, Mg/Fe and Ca/Fe are consistent with the stellar value within 1$\sigma$. However, Ti/Fe is strongly sub-stellar, consistent with multiple studies on WASP-121~b \citep[e.g.,][]{Smith24_W121b, Pelletier2025_wasp121b, Prinoth2025_W121b_4UT}. In WASP-189\,b, Mg/Fe is consistent with the stellar value, whereas Ca/Fe and Ti/Fe are found to be sub-stellar. 

We find a stellar-like Ti/Fe and Ca/Fe in MASCARA-1\,b (Figure~\ref{fig:fetomg}). The Mg/Fe ratio is super-stellar likely due to the super-stellar value retrieved for Mg/H. However, this value is still consistent with stellar value at 2$\sigma$ confidence. The large error on [Mg/Fe]$_\star$ is mostly from the large uncertainty in the stellar Mg abundance (see Appendix \ref{sec:stellar_abundances}). From these, we find that there is no strong indication of nightside cold trapping in MASCARA-1~b.

Figure~\ref{fig:fetomg} shows the retrieved dayside T--P profiles for WASP-189\,b, WASP-121\,b, and MASCARA-1\,b. We also show the nightside GCM predictions (without atmospheric drag) for  WASP-189\,b and WASP-121\,b \citep{Lee2022_GCM_WASP121b_WASP189b}, noting that there is no published prediction for MASCARA-1\,b. The sub-stellar Ti/Fe on WASP-121~b is a strong indication of nightside Ti cold trapping \citep{Smith24_W121b, Pelletier2025_wasp121b, Prinoth2025_W121b_4UT}. The sub-stellar Ca/Fe and Ti/Fe ratios in WASP-189\,b are potentially due to partial nightside cold trapping as the predicted nightside T--P profile crosses the perovskite (CaTiO$_3$) condensation curve.  The stellar Mg/Fe ratios in WASP-189\,b and WASP-121\,b are consistent with the nightside temperature profiles remaining slightly hotter than the Fe condensation curve. 

It is indeed surprising that, although MASCARA-1~b is equally as hot as WASP-189~b in the photosphere (in H \& K bands; $\approx$ 1 - 10$^{-3}$ bar) with higher gravity, cold trapping seems to be less influential. In lieu of an explicit GCM prediction for MASCARA-1\,b, if we approximate the nightside temperature to be $\approx$ 1300 K as reported by \cite{Bell2021}, it is well within reason to expect that the nightside thermal structure of MASCARA-1\,b is cool enough to condense Ti, Ca, and Mg-bearing species. But, we do not find strong evidence for any such depletion in our results.

In this vein, it is also interesting to note that we have retrieved sub-solar Si abundance (by $\sim$2$\sigma$). As discussed in \cite{Woitke2018_GGChem}, the onset of condensation has a significant effect on Si-bearing gases below $\approx$ 1800 K forming gehlenite (Ca$_2$Al$_2$SiO$_7$) and SiO[s] followed by complex silicates. However, given the similar condensation temperatures of Mg and Si, one would expect both Mg and Si abundances to be affected to a similar magnitude. Therefore, it is currently unclear if the root cause of this Mg-enhancement or Si-depletion is due to formation, atmospheric escape, condensation, or a combination of these factors. Additionally, the lack of any MASCARA-1\,b GCM predictions limits our discussion of its nightside thermal structure and its influence on chemistry. For this reason, any future GCMs for MASCARA-1\,b will be greatly valuable to interpret the  refractory ratios from our study.

\section{Conclusions}
\label{sec:conclusions}

We present a comprehensive high-resolution emission spectroscopy analysis of the UHJ, MASCARA-1\,b, using IGRINS and IGRINS-2 on Gemini's South and North Telescopes. Our main findings from this analysis are:

\begin{itemize}

    \item We detect a broad inventory of volatile (H$_2$O at SNR = 4.1, CO at SNR = 5.3, OH at SNR = 5.7) and refractory (Fe\,\textsc{i} at SNR = 9.3, Mg\,\textsc{i} at SNR = 4.4, Ca\,\textsc{i} at SNR = 4.0, Ti\,\textsc{i} at SNR = 4.2) species, providing the most complete atmospheric chemical characterization of MASCARA-1\,b to date.

    \item Chemical equilibrium retrievals yield a solar atmospheric metallicity ([M/H]$_\odot$ $= 0.07^{+0.17}_{-0.13}$ $\approx 1.2\times$ solar), a solar C/O ratio ($= 0.65^{+0.08}_{-0.08}$), and a super-solar refractory enrichment ([$\mathcal{R}$/H]$_\odot$ = $0.40^{+0.23}_{-0.17} \approx 2.5\times$ solar; $\approx 3.8\times$ stellar), corresponding to a moderately enhanced refractory-to-volatile ratio ([$\mathcal{R}/\mathcal{V}$]$_\odot =$ $0.36^{+0.11}_{-0.09}$ $\approx$ 2.3$\times$ solar).
    
    \item Comparison with formation models \citep{Chachan2023} indicates that MASCARA-1\,b most likely accreted material between the soot–H$_2$O or H$_2$O–CO snowlines, with additional scenarios remaining statistically permissible at 2$\sigma$ level.  The relative enhancement of refractory-to-volatile content is consistent with recent IGRINS analyses of two UHJs, WASP-121\,b \citep[][]{Smith24_W121b} and WASP-189\,b \citep{Sanchez26_W189b}.

    \item Species-dependent velocity offsets are broadly consistent with solid-body rotation combined with modest equatorial jets, while minimal $\Delta V_{\mathrm{sys}}$ variations suggest weak large-scale day–night winds, potentially indicative of atmospheric drag. However, the uncertainty on the systemic velocity is to be resolved for a definitive conclusion on presence (or the lack) of atmospheric winds.

    \item We find stellar values for Ti/Fe and Ca/Fe. We obtain a super-stellar value for Mg/Fe, however, this value is consistent with its stellar value at 2$\sigma$ level. Therefore, we find no strong evidence for nightside cold trapping in MASCARA-1~b's atmosphere.

\end{itemize}

Simultaneously constraining volatile and refractory species at high spectral resolution provides a novel pathway to disentangle formation, migration, and atmospheric processing in UHJs. As the Roasting Marshmallows program continues, applying this methodology across a large homogeneous sample of UHJs will lay a foundational work to achieve a population level inference on UHJ formation in the upcoming era of ELTs.


\section*{Acknowledgments}

We thank the NOIRLab staff and the Gemini team (Charles Figura, Leila Alamos, Sunny Stewart, and several others) who helped observe our target on both nights. We also thank Dr.\ Vardan Adibekyan (Instituto de Astrofísica e Ciências do Espaço) and Dr.\ Patrick Young (Arizona State University) for insightful discussions regarding the interpretation of our findings. K.K., M.R.L., and J.L.B.\ acknowledge support for this work from NSF grant AST-2307177. S.P.\ acknowledges support from the Swiss National Science Foundation under grant 51NF40\_205606 within the framework of the National Centre of Competence in Research PlanetS. Based on observations obtained at the international Gemini Observatory, a program of NSF NOIRLab, which is managed by the Association of Universities for Research in Astronomy (AURA) under a cooperative agreement with the U.S. National Science Foundation on behalf of the Gemini Observatory partnership: the U.S. National Science Foundation (United States), National Research Council (Canada), Agencia Nacional de Investigaci\'{o}n y Desarrollo (Chile), Ministerio de Ciencia, Tecnolog\'{i}a e Innovaci\'{o}n (Argentina), Minist\'{e}rio da Ci\^{e}ncia, Tecnologia, Inova\c{c}\~{o}es e Comunica\c{c}\~{o}es (Brazil), and Korea Astronomy and Space Science Institute (Republic of Korea). This work used the Immersion Grating Infrared Spectrometer (IGRINS) that was developed under a collaboration between the University of Texas at Austin and the Korea Astronomy and Space Science Institute (KASI) with the financial support of the Mt. Cuba Astronomical Foundation, of the US National Science Foundation under grants AST-1229522 and AST1702267, of the McDonald Observatory of the University of Texas at Austin, of the Korean GMT Project of KASI, and Gemini Observatory. This work also used the Immersion Grating Infrared Spectrograph 2 (IGRINS-2) developed and built by a collaboration between Korea Astronomy and Space Science Institute (KASI) and the International Gemini Observatory. This research was partly supported by the Korea Astronomy and Space Science Institute (KASI) under the R\&D program (Project No. 2026-1-860-00, Project No. 2026-1-860-01) supervised by the Korea AeroSpace Administration. Finally, we thank the Research Computing at Arizona State University for providing HPC and storage resources that have significantly contributed to the research results reported within this manuscript.

%

\facilities{Gemini:South (IGRINS), Gemini:North (IGRINS-2)}


\software{\textsc{pymultinest} \citep{Buchner2014}, \textsc{astropy} \citep{astropy2022}, \textsc{Numpy} \citep{numpy}, \textsc{Scipy} \citep{scipy}, \textsc{Matplotlib} \citep{matplotlib}, \textsc{fastchem} \citep[][]{Stock2018_fastchem, kitzmann2024}}


\newpage
\appendix

\label{sec:appendix}

\section{A note on adopted Stellar abundances}
\label{sec:stellar_abundances}

We briefly review and summarize the derivation of stellar abundances in the host star MASCARA-1. \citet{Saffe2021} analyzed high-resolution HARPS-N spectra with a signal-to-noise ratio of SNR $\sim$ 560 measured near 5000 \AA. The observed spectra were fitted through an iterative procedure in which specific opacities were computed using ATLAS12 model atmospheres \citep{Kurucz1993_ATLAS12}, which are advanced compared to the classical solar-scaled approach. The final abundance uncertainties include contributions from several factors, such as the line-to-line dispersion and the uncertainties arising from errors in the effective temperature, surface gravity, and microturbulent velocity.

 With regard to chemical species, the C abundance was derived by fitting the C I lines at 5052.17 \AA\ and 5380.34 \AA, since C II lines are too weak or absent in the spectrum of MASCARA-1. The Mg abundance was estimated using the Mg I line at 4702.99 \AA, two lines of the Mg I b triplet (5172.68 \AA\ and 5183.60 \AA), and the strong Mg II line at 4481.23 \AA. Calcium lines commonly observed in early-type stars include Ca I 4226.73 \AA\ and the strong Ca II H and K lines (3933.66 \AA\ and 3968.40 \AA). However, in estimating the Ca abundance, the authors avoided the Ca II 3968.40 \AA\ line because it is heavily blended with the Balmer line H$_{\epsilon}$. Other species, such as Ti and Fe, were analyzed using several dozens of spectral lines. Finally, the obtained stellar abundances of MASCARA-1 are listed in Table \ref{tab:parameters}. The fast rotation ($\varv$ sin\textit{i}$_*$ $\approx$ 101.7 km/sec) of MASCARA-1 poses a significant challenge in obtaining high precision on the stellar elemental abundances. 

\section{Ancillary data}

\begin{figure}
    \centering
    \includegraphics[width=0.7\textwidth]{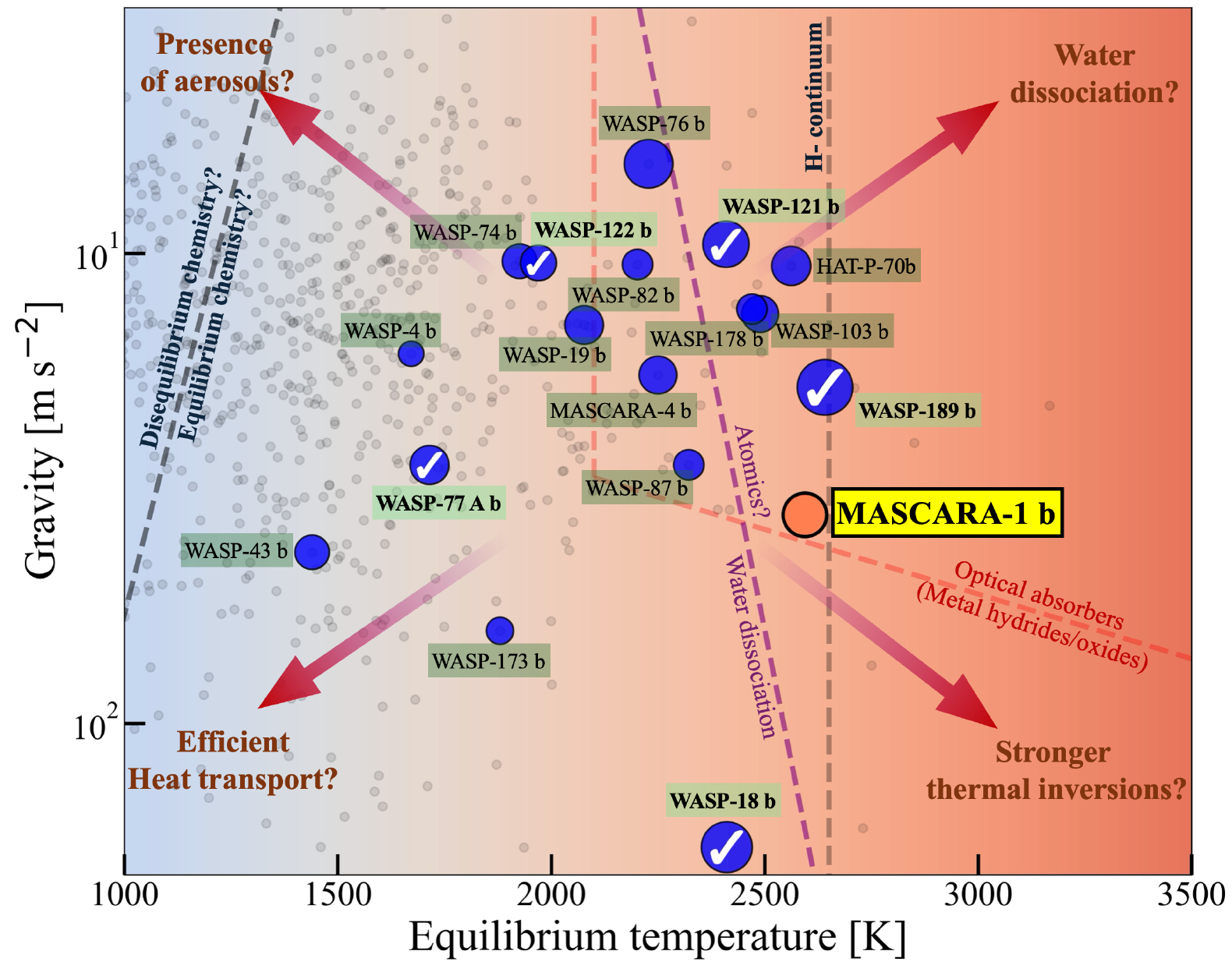}
    \caption{Targets of the Roasting Marshmallows survey (GS-2023B-LP-206; PI: M.~Line) performed using the IGRINS spectrograph on the Gemini-S telescope. In gravity and temperature space, targets of this survey span across key physicochemical transitions enabling inferences on fundamental properties of exoplanet atmospheres. The marker sizes scale with the level of expected thermal emission signal estimated using the emission spectroscopy metric (ESM) relative to WASP-77Ab. Data on transiting exoplanets is shown in the background as unscaled gray dots. At the time of this manuscript's preparation, survey targets with published results have been marked with a white check mark. These are: WASP-77Ab \citep[][]{Line2021_W77Ab}, WASP-18b \citep[][]{brogi23}, WASP-121b \citep[][]{Smith24_W121b}, WASP-122b \citep[][]{Panwar25_W122b}, WASP-189b \citep[][]{Sanchez26_W189b}. MASCARA-1\,b is the second hottest UHJ in this survey and is shown as an orange circle.  This figure is adapted from \citet{Parmentier2018_WASP121b_thermal_dissociation} and the original version made for the Roasting Marshmallows survey.
    }
    \label{fig:survey}
\end{figure}

\begin{figure*}[htbp]
    \centering
    \includegraphics[scale = 0.16]{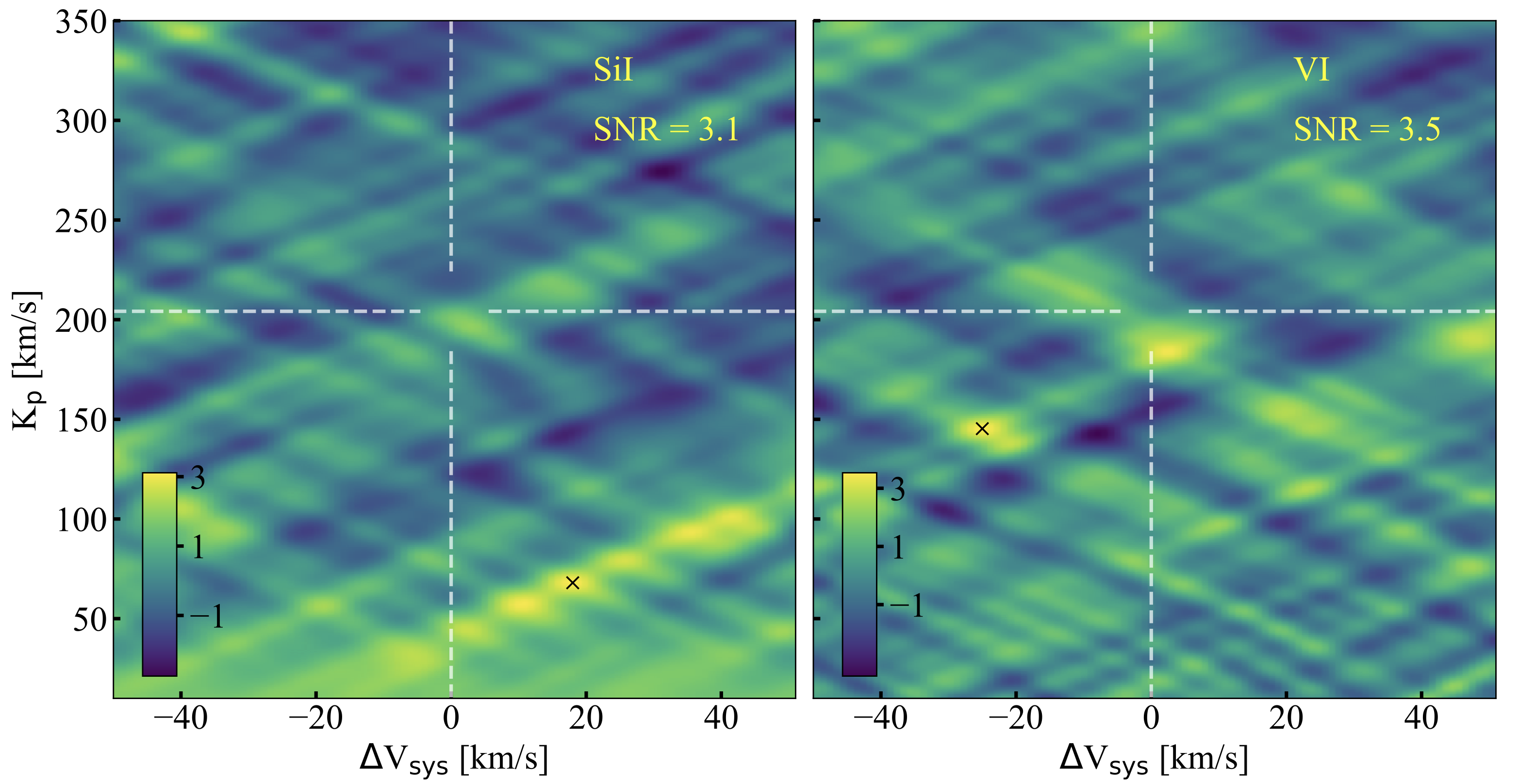}
    \caption{Cross-correlation maps, similar to as shown in Figure \ref{fig:CCmaps}, but for Si\,\textsc{I} and V\,\textsc{I}. Although we do not obtain strong signals from Si\,\textsc{I} and V\,\textsc{I}, we constrain the elemental abundance of Si and V. This is likely due to the weak signal evident near the nominal velocities. However, the peak CC value, shown as the black cross, is significantly offset from the nominal position. Therefore, we interpret elemental abundances of Si and V with caution. Future observations, preferably in the Optical wavelengths where Si~\textsc{i} and V~\textsc{i} have a stronger spectral footprint, are essential to confirm our abundances.}
    \label{fig:nondetections}
\end{figure*}

\begin{figure*}[htbp]
    \centering
    \includegraphics[scale = 0.33]{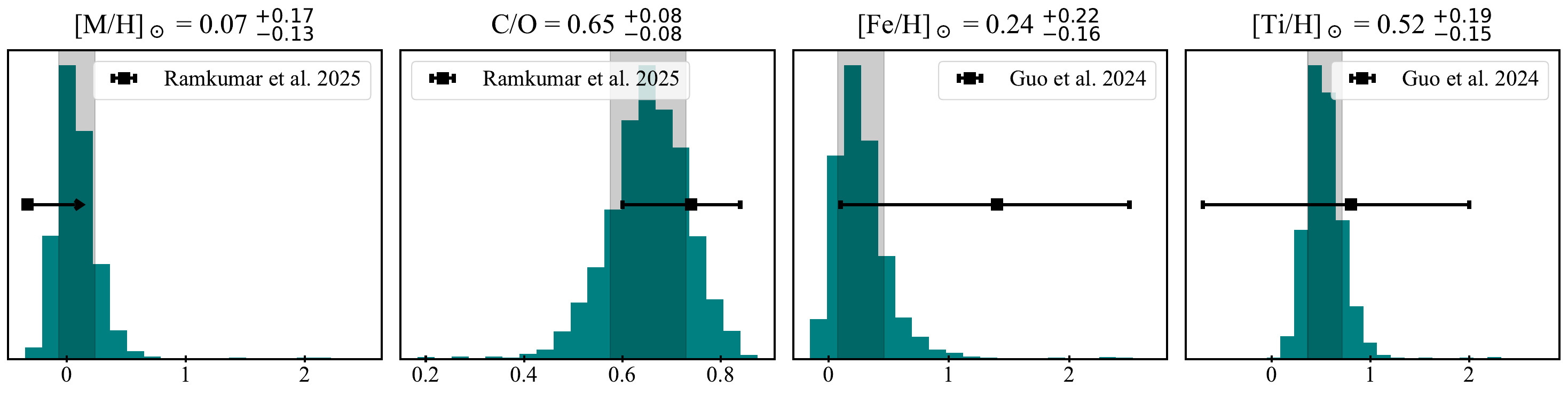}
    \caption{Our retrieved constraints on elemental abundances and derived composition compared with previous works on MASCARA-1\,b by \cite{Ramkumar2025} and \cite{Guo2024}. The shaded regions in all panels shows the errors on our constraints. We find that our fiducial results are consistent with these previous analyses.}
    \label{fig:ratios-compared}
\end{figure*}

\begin{figure}[htbp]
    \centering
    \includegraphics[scale = 0.4]{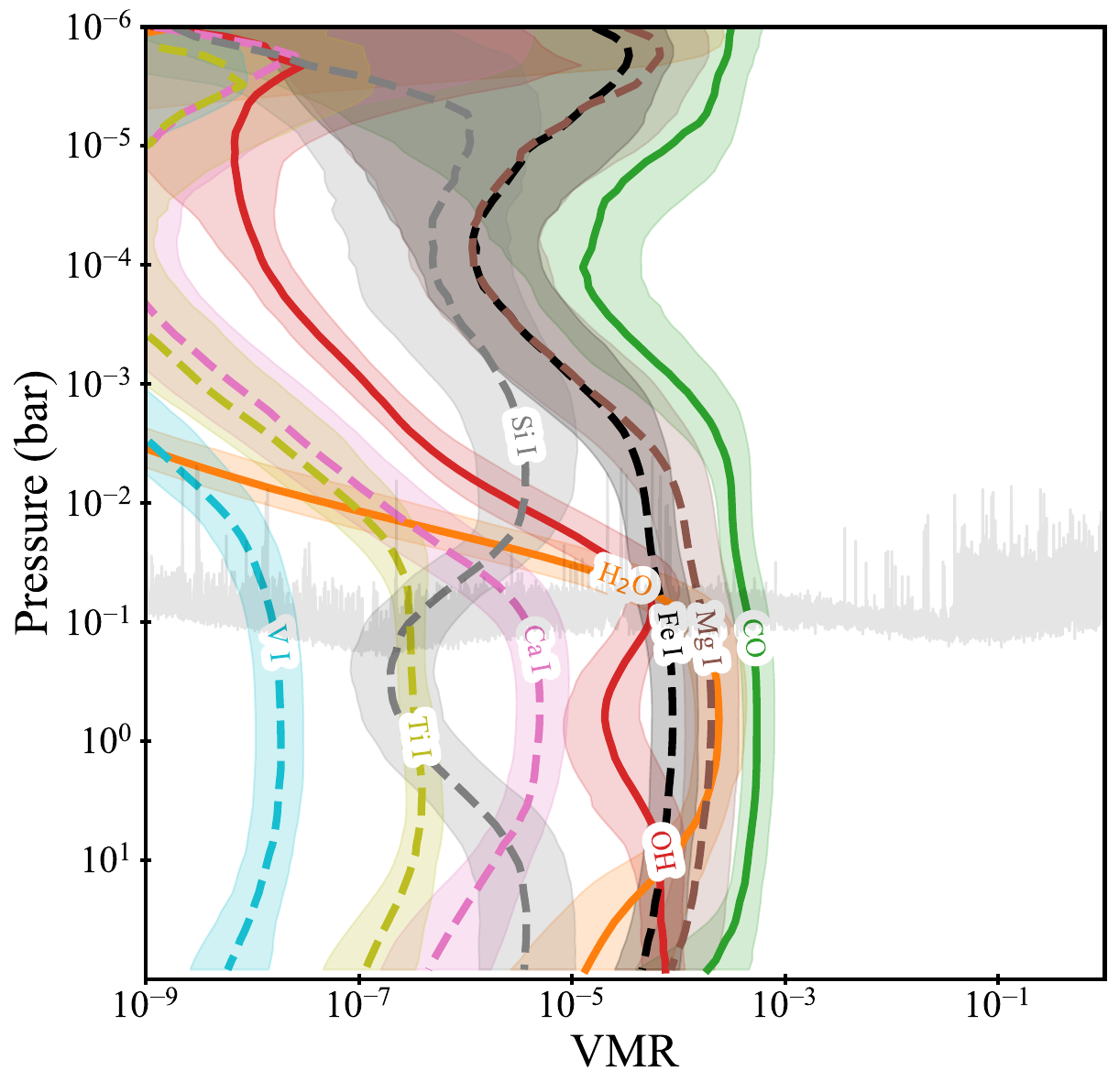}
    \caption{Median volume mixing ratios (VMRs) and their 1$\sigma$ regions obtained using random draws of parameters controlling composition and thermal structure in our retrieval. The gray spectrum in the background denotes the estimated photosphere (where $\tau = $ 2/3). Beyond the photosphere, the precision of VMRs is low as expected.}
    \label{fig:appendix-vmrs}
\end{figure}

\begin{figure}[htbp]
    \centering
    \includegraphics[width=\textwidth]{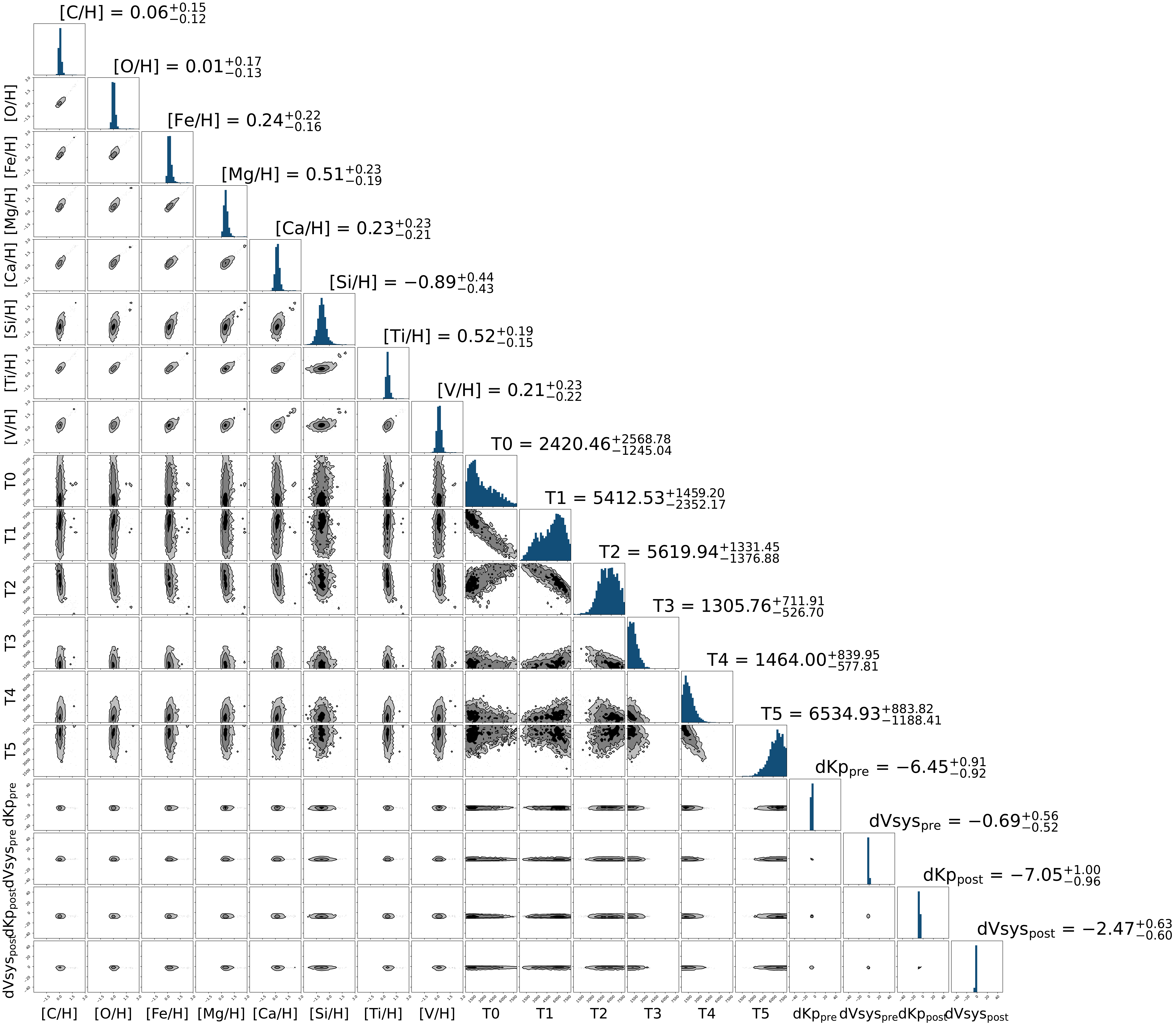}
    \caption{The full corner plot from our retrieval showing the correlations between parameters and their marginalized posterior distributions. We achieved $\approx$ 0.2 dex on all elements except Si. We derive our fiducial constraints on composition using the first eight parameters describing chemistry. We constructed the median thermal structure along with the errors using parameters T0-T5. Overall, we conclude that the atmosphere of MASCARA-1\,b is enriched in refractory elements ($\mathcal{R}$/H) relative to both solar and stellar values.}
    \label{fig:appendix-full-corner}
\end{figure}

In this section, we provide the ancillary tables and additional figures from our analysis. Table \ref{tab:detections} compiles detection results from prior studies alongside our new detections and non-detections. Table \ref{tab:opacity-references} provides references for all opacity sources included in our forward models. Figure \ref{fig:survey} shows the full target list of the Roasting Marshmallows survey, showing targets in surface gravity–equilibrium temperature space along with key physicochemical transition regimes. Figure \ref{fig:nondetections} presents the Kp–$\Delta$V$_\mathrm{sys}$ maps for Si\textsc{I} and V\textsc{I}, which lack confident detections from the CC maps but show a weak signal near the expected Kp–V$_\mathrm{sys}$ velocities. Figure \ref{fig:ratios-compared} benchmarks our fiducial abundance constraints against previous studies. Figure \ref{fig:appendix-vmrs} displays the median VMRs with 1$\sigma$ uncertainties. Figure \ref{fig:appendix-full-corner} shows the full posterior corner plot from our fiducial retrieval.

\begin{table*}[ht!]
\centering
\caption{Previous searches for signals in MASCARA-1\,b's atmosphere}
\label{tab:detections}
\begin{tabular}{lll}
\toprule
Gas & D/ND/T & Reference \\
\midrule
H$_2$O, CO & D & \cite{Holmberg2022, Ramkumar2023, Ramkumar2025} \\
Fe\,\textsc{i} & D & \cite{Ramkumar2023, Ramkumar2025, Guo2024, Scandariato23_M1b} \\
Ti\,\textsc{i} & D & \cite{Guo2024, Scandariato23_M1b} \\
Cr\,\textsc{i} & D & \cite{Scandariato23_M1b} \\
$^{13}$CO, Mg\,\textsc{i}, CO$_2$, OH, HCN, Si\,\textsc{i}, Ca\,\textsc{i} & ND & \cite{Ramkumar2025} \\
\midrule
H$_2$O, CO, OH, Fe\,\textsc{i}, Mg\,\textsc{i}, Ca\,\textsc{i}, Ti\,\textsc{i} & D & This work \\
Cr\,\textsc{i}, Si\,\textsc{i}, V\,\textsc{i}, TiO, VO & ND & This work \\
\bottomrule
\end{tabular}
\end{table*}

\begin{table*}[ht!]
\centering
\caption{References of opacities of gases used in this study. Cross sections for CO and OH were precomputed using HELIOS-K \citep{heliosk2015}. The H$_2$O, TiO, and VO cross-sections were generated as described in \cite{Gharib-Nezhad2021_exoplines}. Atomic species were generated using a custom script that accounts for both natural, pressure, and Doppler broadening (line data from gfall08oct17$^\mathrm{a}$).}
\label{tab:opacity-references}

\begin{tabular*}{0.5\textwidth}{@{\extracolsep{\fill}}cc@{}}
\toprule
\multicolumn{1}{c}{Gas} & \multicolumn{1}{c}{Opacity Reference} \\
\midrule
H$_2$O & \cite{polyansky2018exomol} \\
CO & \cite{li-co-xsecs} \\
OH & \cite{hitemp2010} \\
Fe\,\textsc{i} & \cite{Kurucz_2017_opacities} \\
Mg\,\textsc{i} & \cite{Kurucz_2017_opacities} \\
Ca\,\textsc{i} & \cite{Kurucz_2017_opacities} \\
Si\,\textsc{i} & \cite{Kurucz_2017_opacities} \\
Ti\,\textsc{i} & \cite{Kurucz_2017_opacities} \\
V\,\textsc{i} & \cite{Kurucz_2017_opacities} \\
Cr\,\textsc{i} & \cite{Kurucz_2017_opacities} \\
TiO & \cite{McKemmish2019_TiO} \\
VO & \cite{McKemmish2016_Exomol} \\
CIA (H$_2$-H$_2$ \& H$_2$-He) & \cite{karman2019hitran} \\
H$^-$ (free-free \& bound-free) & \cite{HMBF_HMFF_John_1988} \\
\bottomrule
\multicolumn{2}{c}{\footnotesize $^\mathrm{a}$ \url{http://kurucz.harvard.edu/linelists/gfnew/gfall08oct17.dat}} \\
\end{tabular*}

\end{table*}

\clearpage

\bibliography{m1b_references}{}
\bibliographystyle{aasjournal}



\end{document}